\documentclass[preprint,10pt]{elsarticle}

\usepackage{amssymb}
\usepackage[utf8]{inputenc}
\usepackage[english]{babel}
\usepackage{natbib}
\usepackage{hyperref}
\usepackage{amsmath}
\usepackage{longtable}
\usepackage{geometry}

\biboptions{sort&compress}
\journal{Nuclear Physics A}

\begin{document}

\begin{frontmatter}



\title{Structural Properties and $\alpha$-Decay Chains of Transfermium Nuclei (101$\leq$Z$\leq$110)}


\author[a]{U. K. Singh} \author[a]{R. Sharma}\author[b]{P. K. Sharma}\author[c]{M. Kaushik}\author[a]{S. K. Jain}\author[d]{G. Saxena}
\address[a]{Department of Physics, School of Basic Sciences, Manipal University Jaipur, Jaipur-303007, India}
\address[b]{Govt. Polytechnic College, Rajsamand-313324, India}
\address[c]{S. S. Jain Subodh P. G. College, M. C. A. Institute, Rambagh Circle, Jaipur-302004, India}
\address[d]{Department of Physics (H\&S), Govt. Women Engineering College, Ajmer-305002, India}

\begin{abstract}
Transfermium nuclei (101$\leq$Z$\leq$110) are investigated thoroughly to describe structural properties viz. deformation, radii, shapes, magicity, etc. as well as their probable decay chains. These properties are explored using relativistic mean-field (RMF) approach and compared with other theories along with available experimental data. Neutron numbers N$=$152 and 162 have come forth with a deformed shell gap whereas N$=$184 is ensured as a spherical magic number. The region with N$>$168 bears witness of the phenomenon of shape transition and shape coexistence for all the considered isotopic chains. Experimental $\alpha$-decay half-lives are compared with our theoretical half-lives obtained by using various empirical/semi-empirical formulas. The recent formula proposed by Manjunatha \textit{et al.}, which results best among the considered 10 formulas, is further modified by adding asymmetry dependent terms ($I$ and $I^2$). This modified Manjunatha formula is utilized to predict probable $\alpha$-decay chains that are found in excellent agreement with available experimental data.
\end{abstract}




\end{frontmatter}


\section{Introduction}
\label{intro}
A quest for properties of superheavy elements that are synthesized by cold-fusion or hot-fusion reactions has resulted in very moderate information so far compared to the knowledge gathered for heavy, medium-heavy and light nuclei. Though the growth in experimental facilities has provided new dots upto Z$=$118 on the nuclear landscape but various structural properties are yet to be explored from experimental front. Nevertheless, it has been a long journey by experiments
at GSI \cite{Hofmann2000,Hofmann2011}, RIKEN \cite{Morita2007} and JINR \cite{Oganessian2010,Oganessian2015,Hamilton2013}, and also by various theoretical approaches \cite{Bao2015,Wang2015,Niyti2015,Heenen2015,Santhosh2016,Budaca2016,Liu2017,Zhang2017,TLZhao2018,JPCui2018,saxena121,saxena122} which has comprehended a fairish part of the unknown territory \cite{heenen2015,oganrpp2015,oganpt2015,dull2018,nazar2018,giuliani2019,hofmann2016} and still has many miles to go.\par

For superheavy nuclei to become experimentally accessible, one of the crucial factors is 'stability' which is directly related to their shell structure. Beyond the classical proton and neutron spherical shell gaps at Z$=$82 and N$=$126, several models have made the prediction of various possible shell gaps \cite{Wu2003,Zhang2004,Oganessian2009,Adamian2009,Biswal2014,Brodzinski2013,moller1994,Rutz1996,Cwiok1996}, which are mainly converged to the Z$=$114 and N$=$184. In addition, deformed shell gaps near Z$=$100 and N$=$152 and Z$=$108 and N$=$162 are predicted to occur theoretically \cite{patyk1991,moller1994,cwiok1994,moller2016,stone2019}
as well as experimentally \cite{lazarev1994}. The stability in this region is identified by Hofmann \textit{et al.} \cite{hofmann2001} in a new isotope $^{270}$Ds (N = 160) and by Nishio \textit{et al.} \cite{nishio2010} in another new isotope $^{268}$Hs (N = 160). Further, after the observation of a new nuclide $^{270}$Hs (N = 162) \cite{dvorak2006}, on the basis of Q$_{\alpha}$, the shell gap was confirmed at Z$=$108 and N$=$162 \cite{dvorak2006,oganessian2013}. The systematic of data of spontaneous fission life also demonstrated the stability at N$=$152 with Z$=$98-108 \cite{oganessian2017}.\par

Superheavy nuclei are usually believed to be deformed in shape but so far $^{256}$Fm is the heaviest nucleus for which ground state deformation is available experimentally \cite{iaea}. At the same time, many theoretical treatments have suggested variety of shapes in superheavy nuclei \cite{Heenen2015} and it has been observed that shapes are indeed crucial as the probability of decay is found to depend on the structure of parent and daughter isotopes. Interestingly, the study of decay modes is the leading area on which most of the research of superheavy nuclei is centered and it is veritably found very promising, significant and crucial. \cite{Bao2015,Wang2015,Niyti2015,Heenen2015,Santhosh2016,Budaca2016,Liu2017,Zhang2017,hofmann2016,Oganessian2009}. The most important decay types in the superheavy region are $\alpha$-decay and spontaneous fission (SF) although the possibility of weak decay is also pointed out by various Refs. \cite{hofmann2016,karpov2012,zagrebaev2012,moller2019,hebberger2016,sarriguren2019} and one of our recent work \cite{saxena2020}. The competition between $\alpha$-decay and spontaneous fission (SF)
offers an essential input for the detection of the superheavy nuclei in laboratory and has been employed predominately to predict decay modes or chains \cite{saxena121,saxena122,Bao2015,Wang2015,Niyti2015,Heenen2015,Santhosh2016,Budaca2016,Liu2017,Zhang2017,hofmann2016,Oganessian2009,saxena2020,adamian2020}.\par

From the experimental database \cite{nndc}, information about 15$-$18 isotopes are available for each element in the range 101$\leq$Z$\leq$110, which is reasonably a good number to compare results from theory and thereafter make predictions of unknown territory, expeditiously. With this in view and in continuation to our earlier article \cite{saxena2020}, the transfermium nuclei in the range 101$\leq$Z$\leq$110 are investigated varying neutron number N$\approx$135$-$186 to demonstrate ground-state properties and decay chains for all even and odd isotopes.\par

\section{Theoretical Frameworks}
The calculations in the relativistic mean-field approach \cite{saxena121,Serot1984,Ring1996,Yadav2004,Saxena2017,Saxenaplb2018} have been carried out using the model Lagrangian density with nonlinear terms both for the ${\sigma}$ and ${\omega}$ mesons along with NL3* parametrization \cite{nl3star}. The corresponding Dirac equations for nucleons and Klein-Gordon equations for mesons obtained with the mean-field approximation are solved by the expansion method on the widely used axially deformed Harmonic-Oscillator basis \cite{Geng2003,Gambhir1989}. The quadrupole constrained calculations have been performed for all the nuclei considered here in order to obtain their potential energy surfaces (PESs) and determine the corresponding ground-state deformations \cite{Geng2003,Flocard1973}. For nuclei with an odd number of nucleons, a simple blocking method without breaking the time-reversal symmetry is adopted \cite{Ring1996,Geng2003wt}. In the calculations, for pairing interaction we apply a delta force, i.e., V $=$ -V$_0 \delta(r)$ with the strength V$_0$ $=$ 350 MeV fm$^3$ which has been used in Refs.$~$ \cite{saxena121,Yadav2004,Saxena2017} for the successful description of drip-line nuclei and superheavy nuclei. For further details of these formulations we refer the reader to Refs.$~$\cite{saxena121,Geng2003,Gambhir1989,Singh2013}. \par

As an important fact for this study, the effective mass (Dirac mass) from the relativistic sets is generally very small. In the model used here with cubic and quartic scaler self-interactions, the effective mass (m$^{*}$/m) at saturation is $~$0.60 whereas for realistic Skyrme force like SkM* and optical model analyses it is 0.79 \cite{Ba82,Br85} and 0.83 \cite{Jo87}, respectively. A comprehensive review of different kinds of effective masses within 23 RMF models can be found in ref. \cite{chen2007} or Chapter 4 of ref. \cite{li2008}. Particularly, the shell structure of superheavy nuclei is investigated within various parametrizations of relativistic and nonrelativistic nuclear mean field models \cite{rutz1997} where it has been shown that the spin-orbit splitting shows a principal difference between the linear and the nonlinear sets. The linear sets give too large values of spin-orbit splitting whereas the nonlinear sets are just about right \cite{rutz1997,reinhard1989}. Therefore, even thought relativistic approach uses much smaller values of effective mass but it successfully incorporates the important correlations and is able to reproduce remarkably well the ground state properties including the spin-orbit splittings \cite{latha1994}. The great advantage of the relativistic approach is that it automatically implies a description of spin properties without any extra parameters as required in the non-relativistic case. With these comments, we will probe magicity on the basis of single particle levels which are calculated using non-linear version of parameter set (NL3*) \cite{nl3star} and has been used successfully for describing various ground state properties of superheavy nuclei \cite{saxena121,saxena122}.

\section{Results and discussions}
In the first subsection, we present our results of ground state properties of nuclei 101$\leq$Z$\leq$110 including odd and even nuclei in the range of $^{235-287}$Md, $^{238-288}$No, $^{241-289}$Lr, $^{243-290}$Rf, $^{245-291}$Db, $^{248-292}$Sg, $^{250-293}$Bh, $^{253-294}$Hs, $^{255-295}$Mt, and $^{255-296}$Ds.
All these chains include results up to neutron number N$\approx$184. For all these nuclei, the calculations are done using NL3* \cite{nl3star} parameter of relativistic mean-field theory (RMF) \cite{saxena121,saxena122,Yadav2004,Saxena2017} as explained above. For a comparison, results are compared with Hartree-Fock-Bogoliubov (HFB) mass model with HFB-24 functional \cite{hfbxu}, nuclear mass table with the global mass formula WS4 \cite{ws42014}, recently reported Finite Range Droplet Model (FRDM) calculations \cite{moller2019} along with available experimental data \cite{nndc}. We explore possible new magic numbers, the phenomenon of shape transition and shape co-existence, etc. Competition between $\alpha$-decay and spontaneous fission is analyzed in the second subsection. For $\alpha$-decay, we modify an empirical formula and predict possible $\alpha$-decay chains with their half-lives which are compared with available experimental decay-modes and half-lives.\par

\subsection{Ground State Properties}

\begin{figure}[!htbp]
\centering
\includegraphics[width=0.5\textwidth]{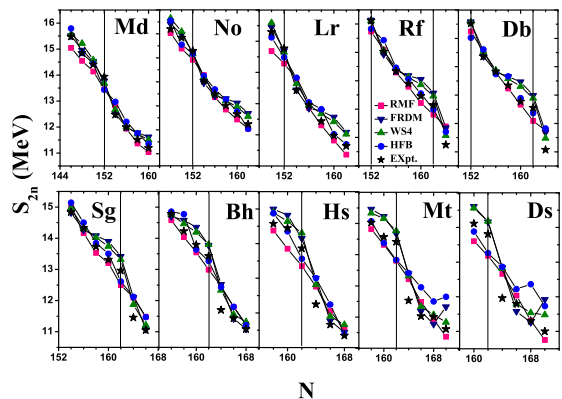}
\caption{(Colour online) Variation of two neutron separation energy (S$_{2n}$) for transfermium nuclei with 101$\leq$Z$\leq$110 and even neutron number N.}
\label{fig1}
\end{figure}

\begin{figure}[!htbp]
\centering
\includegraphics[width=0.5\textwidth]{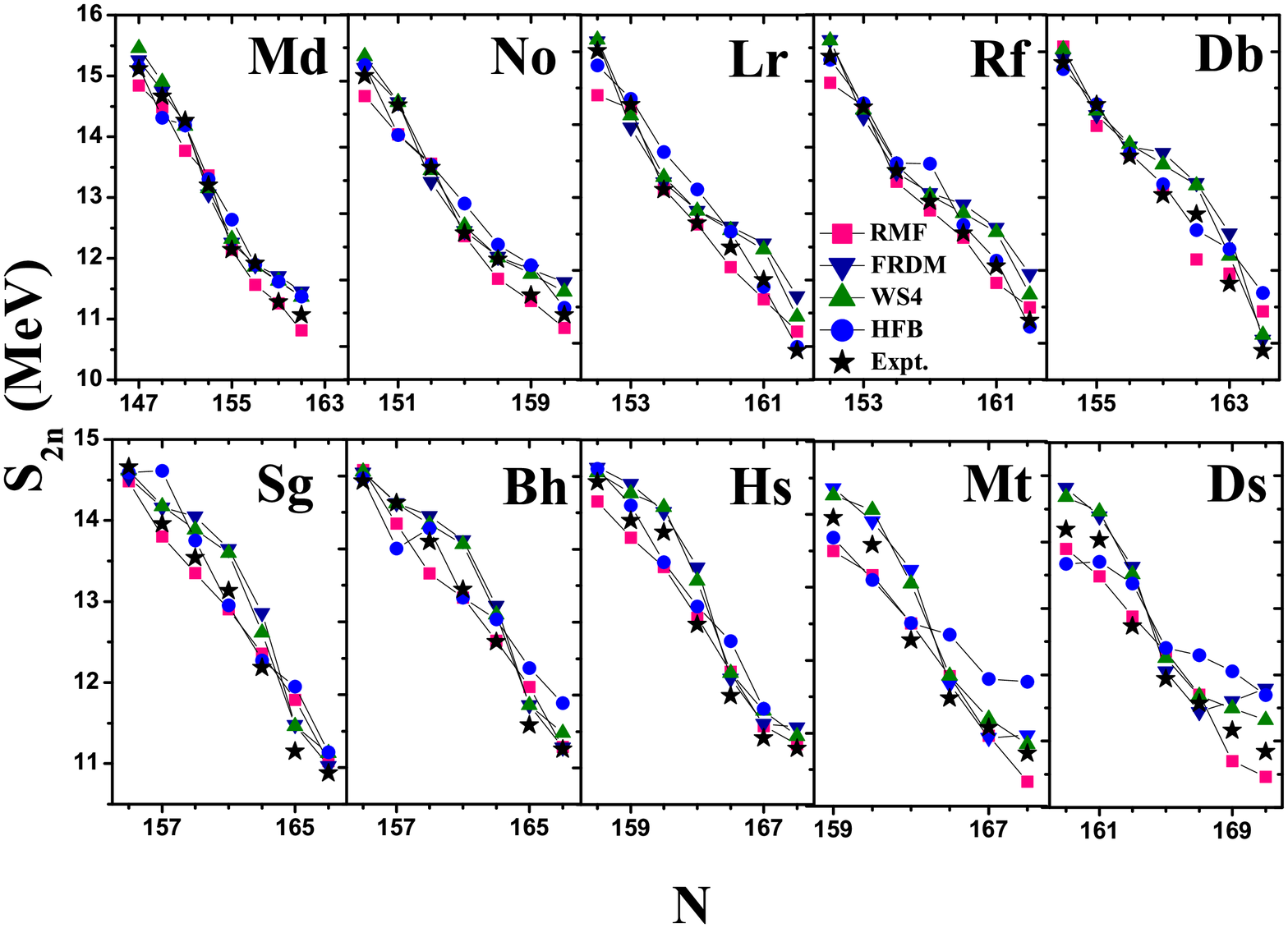}
\caption{(Colour online) Variation of two neutron separation energy (S$_{2n}$) for transfermium nuclei with 101$\leq$Z$\leq$110 and odd neutron number N.}
\label{fig2}
\end{figure}

\begin{table}[!htbp]
\caption{Root mean square error (RMSE) for (S$_{2n}$) for transfermium nuclei with 101$\leq$Z$\leq$110.}
\centering
\resizebox{0.47\textwidth}{!}{%
\begin{tabular}{|c|c|c|c|c|}
\hline
 \multicolumn{1}{|c|}{Nucleus}&
 \multicolumn{4}{c|}{Root mean square error (RMSE)}\\
 \cline{2-5}
&RMF&FRDM&WS4&HFB\\
\hline
Even N&0.3650&	0.3989&	0.3291&	0.4238\\
\hline
Odd N&0.2930&	0.3995&	0.3259&	0.4241\\
\hline
\end{tabular}}
\label{rmses2n}
\end{table}

To demonstrate the predictive power of RMF, we compare two neutron separation energy (S$_{2n}$) with above mentioned theories \cite{moller2019,hfbxu,ws42014}, and experimental data \cite{nndc} for even and odd neutron nuclei separately in Figs. \ref{fig1} and \ref{fig2}. An excellent match with experimental data and also with other theories provide a certificate to RMF theory for the contemplation of these transfermium nuclei. In fact, at various places, RMF theory provides more reasonable agreement with experimental points as compared to other considered theories. To testify this statement, we have calculated root mean square error (RMSE) for the data plotted in Figs. \ref{fig1} and \ref{fig2} which is tabulated in Table \ref{rmses2n}. It is gratifying to note that with the smaller value of RMSE, RMF approach is able to provide an excellent agreement with the experimental values of S$_{2n}$ \cite{nndc} and qualifies to provide a fair description of considered nuclei.\par

\begin{figure}[!htbp]
\centering
\includegraphics[width=0.5\textwidth]{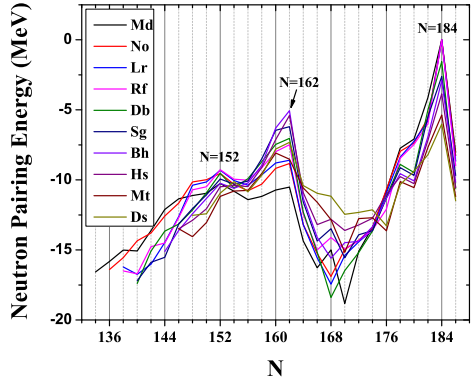}
\caption{(Colour online) Neutron pairing energy contribution for transfermium nuclei with 101$\leq$Z$\leq$110.}
\label{fig3}
\end{figure}

A close look in Fig. \ref{fig1} results in a piece of information related to neutron magicity in this region of the periodic chart. A sharp drop after N$=$152 and N$=$162 comparative to their neighbourhood nuclei manifests magic nature of these neutron number which is in accord with Refs.$~$ \cite{patyk1991,moller1994,cwiok1994,moller2016,stone2019,lazarev1994,hofmann2001,nishio2010}. The experimental systematics of S$_{2n}$ also establish magic nature of N$=$152 and N$=$162. This magic nature can also be verified by systematics of Q$_{\alpha}$ (not shown here). However, RMF model does not show a very steep change at N$=$152 and N$=$162 compared to other models considered here, although the overall calculations are in best agreement as shown in Table \ref{rmses2n}. To show the imprints of magicity from RMF calculations, we have plotted neutron pairing energy contribution for all the considered nuclei in Fig. \ref{fig3}. Vanishing behaviour of pairing energy at N$=$184 clearly indicates strong magicity. In addition, the peaks at N$=$152 and N$=$162, though lower, support magic character of these neutron numbers in agreement with Refs.$~$ \cite{patyk1991,moller1994,cwiok1994,moller2016,stone2019,lazarev1994,hofmann2001,nishio2010}. Since magicity of a nucleus is closely related to its shape, therefore, we look into the shapes of the nuclei having neutrons N$=$152, 162 and 184.\par

\begin{figure}[!htbp]
\centering
\includegraphics[width=0.5\textwidth]{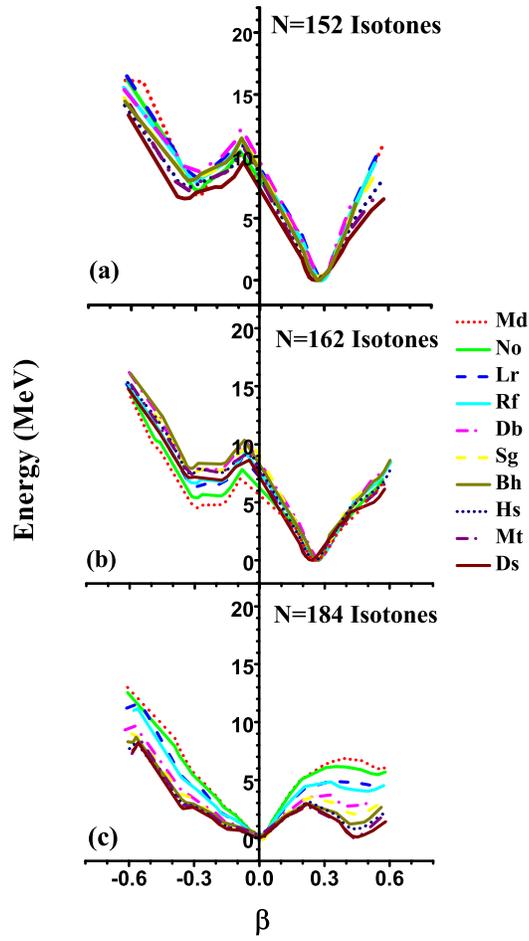}
\caption{(Colour online) Potential energy surfaces (PESs) for isotones of N$=$152, N$=$162 and N$=$184 with 101$\leq$Z$\leq$110.}
\label{fig4}
\end{figure}

\begin{figure}[!htbp]
\centering
\includegraphics[width=0.5\textwidth]{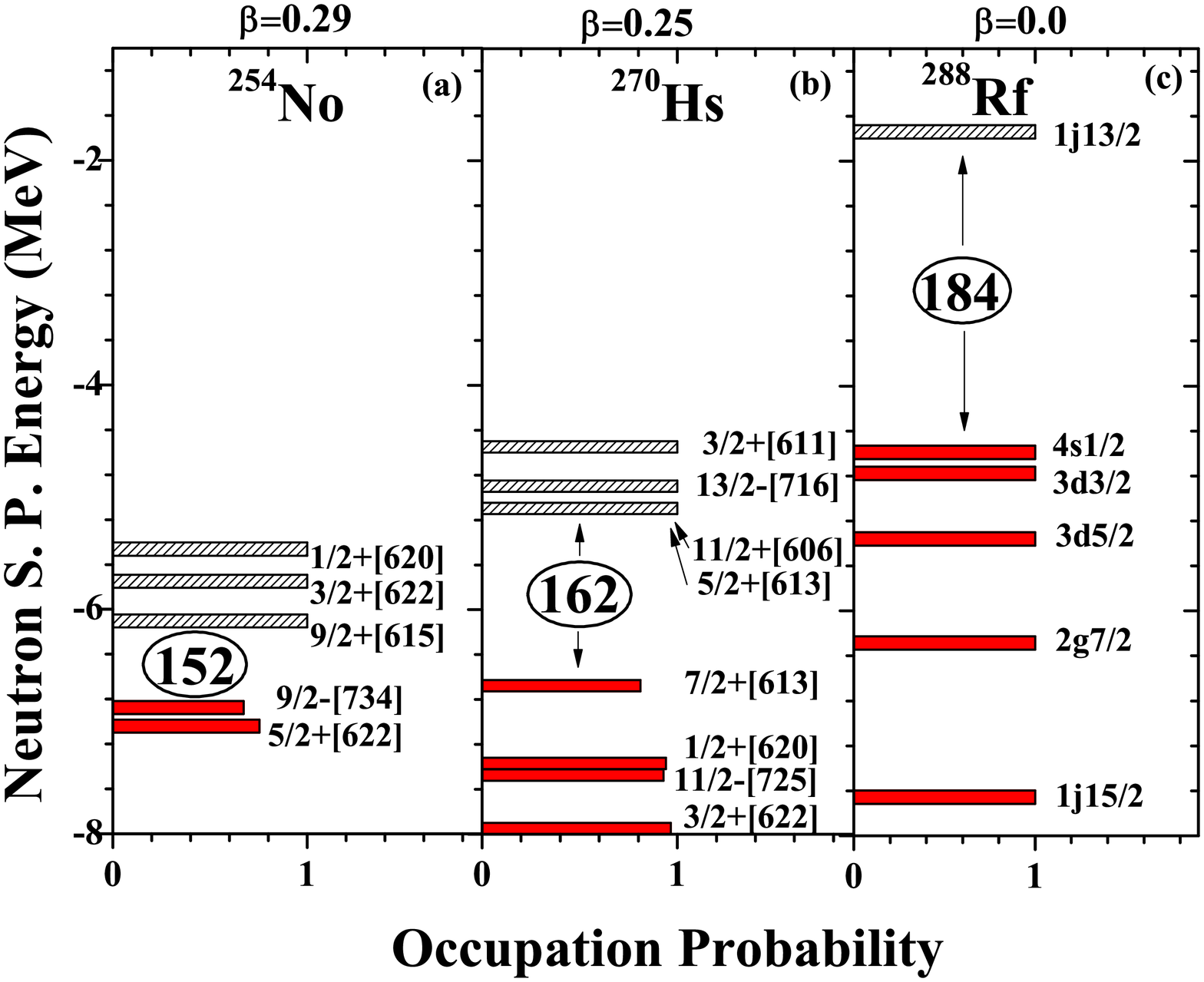}
\caption{(Colour online) Occupancies of neutron single particle states for $^{254}$No, $^{270}$Hs and $^{288}$Rf. States are mentioned in front of their bar. Filled states are shown by red colours whereas empty states are represented by shaded bar.}
\label{fig5}
\end{figure}

The potential energy surfaces (PESs) are plotted in Fig. \ref{fig4} for isotones of N$=$152, N$=$162, and N$=$184 within the range 101$\leq$Z$\leq$110. The energies plotted in Fig. \ref{fig4} are normalized to zero with respect to the lowest values of energy obtained for each isotone. For the case of N$=$184, all the isotones have spherical minimum which is a strong signature of magicity and hence N$=$184 attributes to be a similar magic number as that of the conventional one like N$=$126. However, for almost all the isotones another minimum is observed at quadrupole deformation parameter $\beta$$\approx$0.45. This second minimum is found at a few keV energy for Hs, Mt, and Ds and, therefore, PESs for N$=$184 isotones depict the phenomenon of shape co-existence for $^{292}$Hs, $^{294}$Mt and $^{296}$Ds. On the other side from Figs. \ref{fig4}(a) and (b), only one dominant minimum is observed for all the isotones of N$=$152 and N$=$162 leading to pure deformed nature around $\beta$$\approx$0.3. As a result, no sphericity is detected for any of the nuclei with N$=$152 or N$=$162 which suggests a different character of N$=$152 and 162 compared to N$=$184.\par

To analyze the difference of magicity among N$=$152, N$=$162, and N$=$184, in Fig. \ref{fig5}, we show occupancy of neutron single-particle states for $^{254}$No, $^{270}$Hs and $^{288}$Rf as representative examples of N$=$152, 162 and 184 isotones, respectively. The filled states are represented by red colours whereas the shaded states are empty. The values of quadrupole deformation are mentioned on the top of the graph from which among the three $^{288}$Rf is found with spherical nature having $\beta$$=$0. Therefore, the gap between 6$^{th}$-shell and 1j$^{13/2}$ is found around 2.85 MeV (Fig. \ref{fig5}(c)) which is significant enough to be a shell closure at N$=$184 in this region of the periodic chart. In a similar manner, from Fig. \ref{fig5}(b) the gap between Nilsson orbits 7/2+[613] and 5/2+[613] is ascertained as 1.58 MeV offering a sub-shell characteristic for neutron number N$=$162 in $^{270}$Hs inline with Refs. \cite{patyk1991,moller1994,cwiok1994,moller2016,stone2019,lazarev1994,dvorak2006,oganessian2013}. On the other hand, the gap between Nilsson orbits 9/2$-$[734] and 9/2+[615], responsible for N$=$152, is observed to be very small ($<$1 MeV) in $^{254}$No which reflects weaker magicity in N$=$152 comparative to magic nature of N$=$162 and N$=$184 as also reflected in the peaks of Fig. \ref{fig3}. Hence, magicity at N$=$184 is very much similar as that of the conventional magicity whereas
N$=$162 and N$=$152 are predicted with deformed sub-shell gaps in accord with the work in Refs. \cite{patyk1991,moller1994,cwiok1994,moller2016,stone2019,lazarev1994}. A more detailed investigation of magic nuclei in the superheavy region will be published soon in our upcoming work. \par

\begin{figure}[!htbp]
\centering
\includegraphics[width=0.5\textwidth]{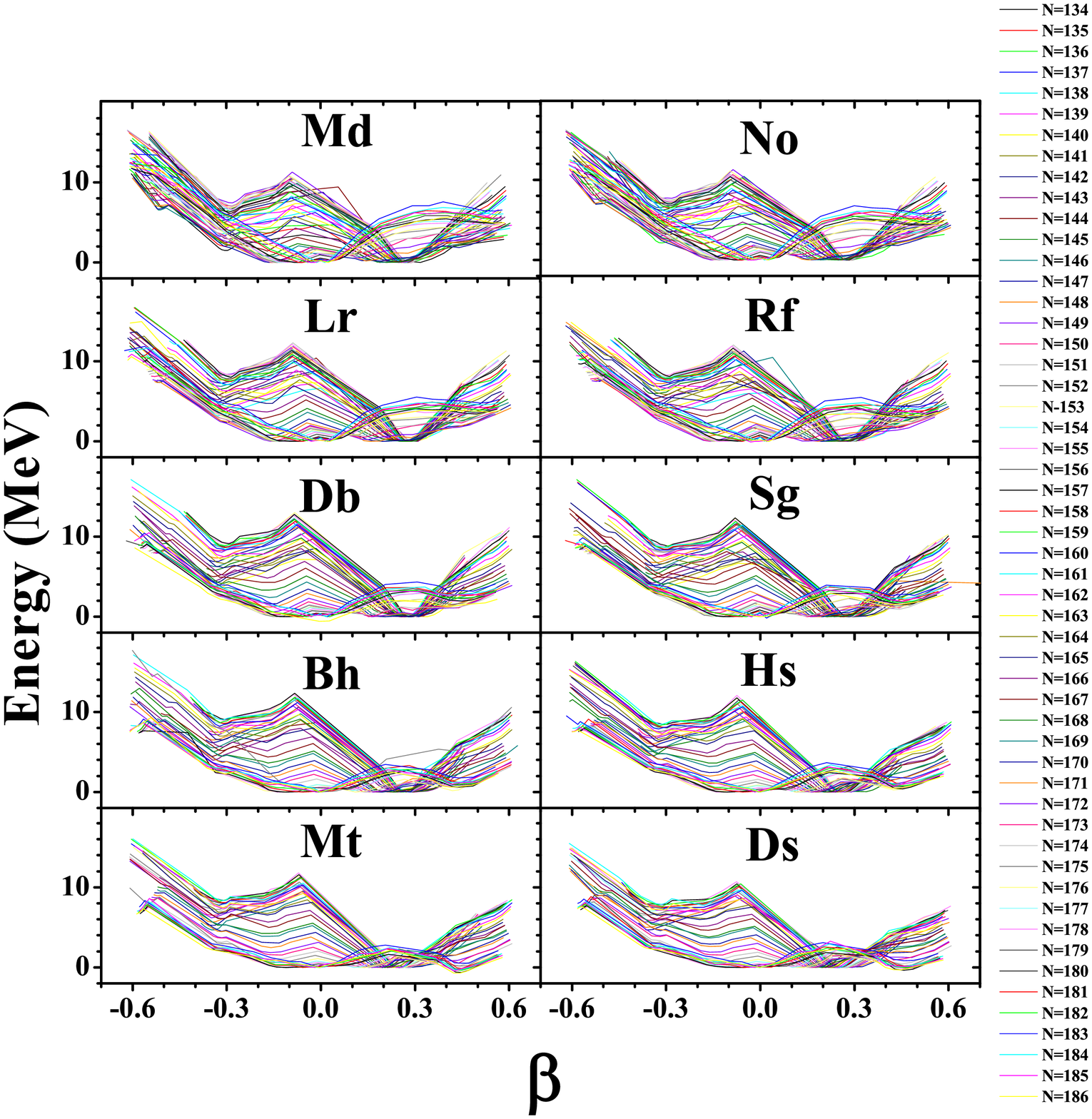}
\caption{(Colour online) Potential energy surfaces for transfermium nuclei with 101$\leq$Z$\leq$110.}
\label{fig6}
\end{figure}

For more insight into the shapes of these nuclei, we have plotted potential energy surfaces of all the nuclei considered here in Fig. \ref{fig6}, by which the evolution of shape and deformation of concern nucleus corresponding to its energy minima can be observed from the variation of binding energies with respect to quadrupole deformation parameter $\beta$. It is worthy to mention here that the full potential energy surface (PES) of superheavy nuclei is of great importance as it allows one to estimate the stability against spontaneous fission and also predicts the most favourable fusion path for the synthesis of these nuclei. It is seen that fission barriers in the region Z$=$100$-$110 are high enough to secure stability with respect to spontaneous fission \cite{cwiok1983}. However, the RMF model systematically predicts lower barriers in this particular superheavy region \cite{burvenich2004}. For this particular study, we have not calculated fission barrier and restricted our calculations upto axially symmetric shapes. The inclusion of triaxial degree of freedom and then the calculations of fission barrier is left for further study.\par

The energies plotted in Fig. \ref{fig6} are normalized to zero with respect to the lowest values of energy obtained for each isotope. The development of a variety of shapes is observed in these transfermium nuclei while moving from the neutron deficient to the neutron-rich side. In majority, the nuclei are found with prolate dominant shape ranging neutron number N$\approx$135-168 as can be seen from all the panels of Fig. \ref{fig6}. For N$\geq$168 the nuclei are switched to oblate shape as a dominant shape, however, interestingly, some of the nuclei are found with the coexisting oblate and spherical minima and hence referred as the potential candidates of shape coexistence. Near N$=$184, all the nuclei are found predominantly with spherical shape as also shown in Fig. \ref{fig4}.\par

The numerical values of deformation along with other ground-state properties viz. binding energy, charge radius, neutron radius, matter radius, etc. for all considered transfermium isotopes are tabulated in the Appendix. From the tables in Appendix, prolate shape for nuclei within 135$\leq$N$\leq$168 is clearly visible. Shape transition from prolate to oblate shape for N$\geq$168 is found for all elements between 101$\leq$Z$\leq$110. Shape transition and its correlation with charge radii can be affirmed by a change around at N$\approx$168 establishing the correlation of shape with charge radius. This kind of change in the charge radius is attributed to the shape transition which is recently observed and demonstrated for the isotopes of Z$=$122 and 128 \cite{tasleem2020}. Another change in the charge radii is found while the oblate shapes switch to spherical shapes at N$=$184 (Kindly refer to Tables \ref{gs-properties-Md-Lo}-\ref{gs-properties-Mt}.

\subsection{$\alpha$-Decay Chains}

We have used various empirical/semi-empirical formulas of $\alpha$-decay to identify the best possible formula for the prediction of decay-modes, particularly for the range 101$\leq$Z$\leq$110. The formulas used in this article are Viola and Seaborg (VSS) \cite{vss1989}, Brown \cite{brown1992}, Royer \cite{royer2000}, Sobiczewski \cite{sobiczewski2005}, universal decay law (UDL) introduced by Qi \textit{et al.} \cite{qi2009}, modified Royer formula given by Akrawy \textit{et al.} \cite{akrawy2017}, modified UDL formula \cite{akrawy2019}, MSLB formula \cite{akrawyprc2019}, modified Viola-Seaborg formula \cite{akrawyprc2019} and Manjunatha present formula \cite{manjunatha2019}. With the use of these formulas, we have calculated $\alpha$-decay half-lives using our calculated values of Q$_{\alpha}$ and compared the half-lives with available experimental data \cite{nndc}. The root mean square error (RMSE) for almost 60 nuclei in the range 101$\leq$Z$\leq$110 by all the above formula are shown in the Table \ref{rmsealpha}. Commendably, the formula provided by Manjunatha \textit{et al.} \cite{manjunatha2019} has provided best results for our considered isotopes and the value of RMSE ($=$2.14) is found smallest among the above considered formulas.\par

\begin{table}[!htbp]
\caption{Root mean square error (RMSE) for $\alpha$-decay half-lives calculated for transfermium nuclei with 101$\leq$Z$\leq$110.}
\centering
\resizebox{0.57\textwidth}{!}{%
\begin{tabular}{|l|c|}
\hline
Formula & RMSE \\
\hline
VSS 1989 \cite{vss1989}                  & 3.54       \\
Brown 1992 \cite{brown1992}              & 2.43       \\
Royer 2000 \cite{royer2000}              & 3.38       \\
Sobiczewski 2005 \cite{sobiczewski2005}  & 3.44       \\
UDL 2009 \cite{qi2009}                   & 3.09       \\
Akrawy 2017 \cite{akrawy2017}          &   3.41                   \\
MUDL 2019 \cite{akrawy2019}            &   4.07                   \\
MSLB 2019 \cite{akrawyprc2019}         &   3.21                   \\
MVS 2019 \cite{akrawyprc2019}          &   3.82                   \\
Manjunatha 2019 \cite{manjunatha2019}  &   2.14                   \\
Modified Manjunatha formula (MMF)-2020 (in this work) & 1.79\\
 \hline
\end{tabular}}
\label{rmsealpha}
\end{table}

The formula proposed by Manjunatha \textit{et al.} is given below \cite{manjunatha2019}:
 \begin{eqnarray}
 log_{10}T_{\alpha}(sec) &=&  8.0393535 (Z_d^{0.4}/\sqrt{Q_{\alpha}})^2 -  18.389095 (Z_d^{0.4}/\sqrt{Q_{\alpha}}) + 2.9699988
 \label{alpha}
\end{eqnarray}

In the present work, for a more accurate prediction of $\alpha$-decay chains, the formula of Manjunatha \textit{et al.} \cite{manjunatha2019} is modified by introducing asymmetry dependent terms ($I$ and $I^2$) which are linearly related to the logarithm of $\alpha$-decay half-lives \cite{akrawyprc2019}. After introducing these terms the formula is further fitted using experimental half-lives of 366 nuclei in the range 82$\leq$Z$\leq$117 and the coefficients are deduced for even and odd cases separately. We have also attempted the fitting using only $I$ or $I^2$ term but the RMSE values are found slightly higher side i.e. 1.8362 and 1.8150, respectively comparative to the formula with $I$ and $I^2$ terms both (RMSE$=$1.7935).

The modified Manjunatha formula (MMF-2020) is given by:

 \begin{equation}
 log_{10}T_{\alpha}(sec) =  a (Z_d^{0.4}/\sqrt{Q_{\alpha}})^2 + b (Z_d^{0.4}/\sqrt{Q_{\alpha}}) + c + dI + eI^2
 \label{modify-alpha}
\end{equation}

here, Z$_d$ is atomic number of daughter nucleus and coefficients are given in Table \ref{coeff}:

\begin{table}[!htbp]
\caption{The coefficients of modified Manjunatha formula (MMF-2020).}
\centering
\resizebox{0.7\textwidth}{!}{%
\begin{tabular}{|l|c|c|c|c|c|}
\hline
Sets       &    a      &         b      &     c   &         d&      e\\
\hline
Even-even  &  -5.57326 &	48.18490	&-80.35736&	5.12241	 &   27.43124  \\
Even-odd   &   5.91103 &    -5.40389	&-16.24855&	-23.34471&	135.61109 \\
Odd-even   &   2.20340 &    12.14202	&-40.09277&	21.60729&	-13.56287 \\
Odd-odd    &   -7.50692&	49.99578	&-70.73322&	-48.97284&	215.35032  \\
\hline
\end{tabular}}
\label{coeff}
\end{table}

After using I-dependence, the modified Manjunatha formula (MMF$-$2020) is applied for the same set of nuclei and has provided ameliorated results as can be seen from the Table \ref{rmsealpha} in which the RMSE value has been improved from 2.14 to 1.79. Henceforth, in this article for the accurate prediction of $\alpha$-decay half-lives and construction of $\alpha$-decay chains we employ the modified Manjunatha formula introduced above. It is worthy to note here that such kind of empirical formulas are fairly dependent on Q$\alpha$ and therefore it is very important to compare Q$\alpha$ of our theoretical formalism (RMF) with the available experimental data. In this regard, we have calculated RMSE for the values of Q$\alpha$ for the range 101$\leq$Z$\leq$110 obtained by using RMF along with some other theories. viz. FRDM, WS4 and HFB. The values of RMSE are found 0.4788, 0.4792, 0.2671 and 0.4843, respectively, strengthening the outcomes based of modified Manjunatha formula (MMF$-$2020) using Q$\alpha$ from RMF theory. Another significant point is the sensitivity of half life on values of Q$\alpha$. The Manjunatha formula \cite{manjunatha2019} has been proved to be reasonably less sensitive comparative to various other empirical formula by calculating the change of alpha decay half-life (in s) by changing Q$\alpha$-value. To check the sensitivity of modified Manjunatha formula (MMF$-$2020), we have also calculated ROC (rate of change) of half-life with respect to Q$\alpha$-value using the following equation:
      \begin{equation}
        ROC= \left[\frac{T_{Q_\alpha}}{T_{Q_{\alpha} + 0.05 MeV}}-1\right]\times100\%
      \end{equation}
 The values of ROC are calculated for the step change of 0.05 MeV in Q$\alpha$-values upto 0.6 MeV. The average of all the ROC from Manjunatha formula and modified Manjunatha formula are found to be 184.74 and 97.44, respectively, which fortifies the lesser sensitivity of the proposed formula. With the above remarks, MMF-2020, characterizes small sensitivity on Q$\alpha$ and combining with lesser RMSE in Q$\alpha$ values from RMF, qualifies to predict the half lives and decay chain in the unknown territory of superheavy region. However, a deeper study on the sensitivity of proposed empirical formula and the variance of the Q$\alpha$-values is required separately. \par

 In the superheavy region the dominant decay mode is $\alpha$, however, it always competes with spontaneous fission (SF), therefore, to construct decay chain in this region, $\alpha$-decay half-lives are compared with spontaneous fission half-lives using the formula given by Karpov \textit{et al.} \cite{karpov2012} and modified with new even-odd corrections by Silisteanu \textit{et al.} \cite{Silisteanu2015,anghel2017}. It is worthy to mention here that this modified formula \cite{Silisteanu2015} of SF is successfully applied for the isotopes of elements with Z$=$104$-$112 \cite{Silisteanu2015,anghel2017}.

\begin{eqnarray}
        log_{10}T_{SF}(sec) & = & 1146.44 - 75.3153Z^2/A + 1.63792(Z^2/A)^2\nonumber\\
& & - 0.0119827 (Z^2/A)^3 \nonumber\\
& &+ B_f  (7.23613 - 0.0947022Z^2/A)\nonumber\\
& & + h_{e-o}
\label{TSF}
\end{eqnarray}
Here $B_f$ is the fission barrier, which is calculated as a sum of the liquid-drop barrier $B_f(LDM)$ and the ground state
shell correction $\delta U(g.s.)$, i.e. $B_f$ = $B_f (LDM)$ + $\delta U(g.s.)$ \cite{karpov2012}. For our calculation, we take fission barrier $B_f$ directly from the Ref.$~$ \cite{moller2015}. Here, $h_{e-o}$$=$0 for (Z$=e$, N$=e$), 2.007 for (Z$=e$, N$=o$), 2.822 for (Z$=o$, N$=e$), and 3.357 for (Z$=o$, N$=o$).\par

Therefore, with the use of Equations. \ref{modify-alpha} and \ref{TSF}, we investigate the contest between $\alpha$-decay and spontaneous fission, and predict the most probable decay modes for all the considered isotopes of the region between 101$\leq$Z$\leq$110 and afterwards construct the possible decay chains along with their $\alpha$-decay/SF half-lives. These potential decay chains are shown in Tables \ref{Ds-chain} and \ref{Mt-chain} in which $\alpha$-decay half-lives and probable decay modes are compared with available experimental data \cite{nndc}. The RMF theory together with the modified Manjunatha formula indeed provides an excellent agreement with the experimental half-lives and decay modes which can be seen in Tables \ref{Ds-chain} and \ref{Mt-chain}, and, therefore, capable to anticipate probable decay chains efficaciously. The variance in decay modes is found only in a few cases in Tables \ref{Ds-chain} and \ref{Mt-chain} which may be due to the fact that the spontaneous fission half-lives are calculated by using the formula of Silisteanu \textit{et al.} \cite{Silisteanu2015} that is fitted by experimental data upto 2015. A new fit with latest experimental data in the formula of spontaneous fission formula may lead to better agreement which is one of the objective of our upcoming work \cite{gauravjpg2020}. Tables \ref{Ds-chain} and \ref{Mt-chain} show the decay chains starting from Ds and Mt for odd and even A nuclei. This comprehensive study of all decay chains completely covers the transfermium region between 101$\leq$Z$\leq$110.

\begin{table}[!htbp]
\caption{$\alpha$-Decay chains of even and odd isotopes of transfermium nuclei starting from Ds and Mt. $\alpha$-decay half-lives are evaluated by using modified Manjunatha formula (MMF-2020) introduced in the present work. Experimental data are taken from Ref. \cite{nndc}.}
\centering
\resizebox{1.0\textwidth}{!}{%
\begin{tabular}{|c|c|c|c|c|c|c|c|c|c|c|c|c|}
\cline{1-6}\cline{8-13}
 \multicolumn{1}{|c}{Nucleus}&
 \multicolumn{3}{|c}{log$_{10}$T$_{1/2}$(s)}&
\multicolumn{2}{|c}{Decay Modes}&
 \multicolumn{1}{|c|}{}&
  \multicolumn{1}{|c}{Nucleus}&
 \multicolumn{3}{|c}{log$_{10}$T$_{1/2}$(s)}&
  \multicolumn{2}{|c|}{Decay Modes}\\
  \cline{2-6}\cline{9-13}
 \multicolumn{1}{|c}{}&
 \multicolumn{1}{|c}{$\alpha$(MMF)}& \multicolumn{1}{|c}{SF}&
  \multicolumn{1}{|c}{Expt.}&\multicolumn{1}{|c}{Predicted}&\multicolumn{1}{|c|}{Expt.}&&
\multicolumn{1}{|c}{}&
 \multicolumn{1}{|c}{$\alpha$(MMF)}& \multicolumn{1}{|c}{SF}&
  \multicolumn{1}{|c|}{Expt.}&\multicolumn{1}{c}{Predicted}&\multicolumn{1}{|c|}{Expt.}\\
\cline{1-6}\cline{8-13}
$^{281}$Ds &  2.28   &  4.15       & 0.98 &  $\alpha$  &  $\alpha$ &&  $^{270}$Ds  &-4.19   & 1.54  & -4.00   &$\alpha$   &$\alpha$   \\
$^{277}$Hs &  3.71   &  5.05       & -2.52 &  $\alpha$  &  $\alpha$&&  $^{266}$Hs  &-2.71   & 2.27  & -2.64   &$\alpha$   &$\alpha$   \\
$^{273}$Sg &  3.64   &  6.08       &     &  $\alpha$  &            &&  $^{262}$Sg  &0.34    & 2.56  &  -2.16 & $\alpha$  & SF/$\alpha$  \\
$^{269}$Rf &  7.61   &  8.45       &     &  $\alpha$  &            &&  $^{258}$Rf  &2.49    & 3.26  &  -1.83 & $\alpha$  & SF/$\alpha$  \\
$^{265}$No &  10.03  &  10.06      &     &  $\alpha$  &            &&  $^{254}$No  &3.34    & 8.75  &  1.71  &  $\alpha$ &  $\alpha$ \\
\cline{1-6}\cline{8-13} 
$^{280}$Ds &  1.32  &  1.38      &     &  $\alpha$  &               && $^{269}$Ds   &-2.88   & 1.84  &-3.75   &$\alpha$   &$\alpha$   \\
$^{276}$Hs &  1.16  &  2.28      &     &  $\alpha$  &               && $^{265}$Hs   &-2.19   & 4.05  &-2.72   &$\alpha$   &$\alpha$   \\
$^{272}$Sg &  2.71  &  5.31      &     &  $\alpha$  &               && $^{261}$Sg    &0.27   & 4.23  & -0.75 & $\alpha$  & $\alpha$  \\
$^{268}$Rf &  6.73  &  7.58      &     &  $\alpha$  &               && $^{257}$Rf    &2.15   & 6.19  & 0.64  &  $\alpha$ &  $\alpha$ \\
$^{264}$No &  9.26  &  7.52      &     &  SF        &               && $^{253}$No    &2.58   & 10.99 & 1.99  & $\alpha$  & $\alpha$  \\
\cline{1-6}\cline{8-13}
$^{279}$Ds &  1.41   & 3.01  &  -0.74    &  $\alpha$  &  $\alpha$    && $^{268}$Ds   &-4.75   & -1.76 &         &$\alpha$   &   \\
$^{275}$Hs &  1.62   & 5.84  &  -0.82    &  $\alpha$  &  $\alpha$    && $^{264}$Hs   &-3.17   & 1.42  &-3.10    &$\alpha$   &$\alpha$   \\
$^{271}$Sg &  3.16   & 8.82  &  2.16     &  $\alpha$  & $\alpha$/SF  && $^{260}$Sg   &-0.75   & 1.85  & -2.31  & $\alpha$  & $\alpha$/SF  \\
$^{267}$Rf &  5.53   & 10.61 &           &  $\alpha$  &              && $^{256}$Rf   &1.09    & 4.69  & -2.18  & $\alpha$  & SF/$\alpha$  \\
$^{263}$No &  11.65  & 8.80  &           &  SF        &              && $^{252}$No   &2.54    &  7.22 & 0.39   &  $\alpha$ &  $\alpha$/SF \\
\cline{1-6}\cline{8-13}
$^{278}$Ds &  -0.54  &  0.64       &    &  $\alpha$  &               && $^{267}$Ds   &-3.77   & 0.01 &-5.55     &$\alpha$   &$\alpha$   \\
$^{274}$Hs &  -0.73  &  4.77       &    &  $\alpha$  &               && $^{263}$Hs   &-2.34   & 2.84 &-3.13     &$\alpha$   &$\alpha$   \\
$^{270}$Sg &  2.93   &  7.98       &    &  $\alpha$  &               && $^{259}$Sg   &-0.72   & 3.56 & -0.54   & $\alpha$  & $\alpha$  \\
$^{266}$Rf &  6.21   &  8.27       &    &  $\alpha$  &               && $^{255}$Rf   &0.54    & 6.36 & 0.23   &  $\alpha$ &  $\alpha$/SF \\
$^{262}$No &  8.68   &  5.91       &-2.30&  SF        &   SF         && $^{251}$No   &2.66    & 8.08 & -0.10   & $\alpha$  & $\alpha$  \\
\cline{1-6}\cline{8-13}
$^{277}$Ds &  -0.23   & 3.98    & -2.39    &  $\alpha$  &  $\alpha$       && $^{266}$Ds   &-5.09   &-2.58  & &$\alpha$   &  \\
$^{273}$Hs &  -0.00014& 8.06    & -0.12    &  $\alpha$  &  $\alpha$       && $^{263}$Hs   &-3.71   &0.22   & &$\alpha$   &  \\
$^{269}$Sg &  1.97    & 10.94   & 2.27     &  $\alpha$  &  $\alpha$       && $^{258}$Sg   &-1.89   &1.66   & & $\alpha$  &  \\
$^{265}$Rf &  7.89    & 9.41    & 1.78     &  $\alpha$  &  SF             && $^{254}$Rf   &0.46    &2.89   & &  $\alpha$ & \\
$^{261}$No &  8.76    & 8.07    &    &  SF        &                       && $^{250}$No   &1.23    &4.38   & & $\alpha$  &  \\
\cline{1-6}\cline{8-13}
$^{276}$Ds &  -2.07  &  2.99      &   &  $\alpha$  &                   &&  $^{265}$Ds   &-3.82   &-0.75  &   &$\alpha$   &   \\
$^{272}$Hs &  -1.51  &  6.98      &   &  $\alpha$  &                   &&  $^{261}$Hs   &-2.80   &1.31   &  & $\alpha$  &   \\
$^{268}$Sg &  2.90  &   8.61      &   &  $\alpha$  &                   &&  $^{257}$Sg   &-1.97   &2.72   &  & $\alpha$  &   \\
$^{264}$Rf &  5.71  &   5.47      &   &  SF       &                    &&  $^{253}$Rf   &0.68    &3.56   & -4.32   & $\alpha$  &$\alpha$    \\
$^{260}$No &  7.91  &   4.57      &-0.97   &  SF        &   SF         &&  $^{249}$No   &0.91    &5.37   & -5.70   & $\alpha$  &$\alpha$    \\
\cline{1-6}\cline{8-13}
$^{275}$Ds &  -1.56  & 6.03         &    &  $\alpha$  &                      && $^{264}$Ds    &-5.70   &-3.20   &   &$\alpha$   &   \\
$^{271}$Hs &  -1.89  & 9.33         &    &  $\alpha$  &                      && $^{260}$Hs    &-4.37   &-1.68   &   &$\alpha$   &   \\
$^{267}$Sg &  4.50   & 9.64         &    &  $\alpha$  &                      && $^{256}$Sg    &-2.31   &-0.75   &   &$\alpha$   &   \\
$^{263}$Rf &  5.50   & 6.57         & 2.78   &  $\alpha$  & SF               && $^{252}$Rf    &-0.83   &-0.10   &   &$\alpha$   &   \\
$^{259}$No &  8.20   & 6.47         & 3.54   &  SF        & $\alpha$         && $^{248}$No    &0.56    &1.87    &   &$\alpha$   &   \\
\cline{1-6}\cline{8-13}
$^{274}$Ds & -3.16  &4.93   &         &  $\alpha$  &              &&   $^{280}$Mt &  3.72  &   5.20      &   &  $\alpha$  &    \\
$^{270}$Hs & -1.94  &6.69   & 1.34    &  $\alpha$  &  $\alpha$    &&   $^{276}$Bh &  4.34  &   5.67      &   &  $\alpha$  &    \\
$^{266}$Sg & 2.74   &6.09   & 1.32    &  $\alpha$  &  SF/$\alpha$ &&   $^{272}$Db &  5.75  &   7.66      &   &  $\alpha$  &    \\
$^{262}$Rf & 4.67   &4.20   & 0.36    &  SF        & SF           &&   $^{268}$Lr &  8.41  &   9.95      &   &  $\alpha$  &    \\
$^{258}$No & 6.41   &4.06   & -2.92   &  SF        & SF           &&   $^{264}$Md &  9.26   & 11.30      &    & $\alpha$  &    \\
\cline{1-6}\cline{8-13}
$^{273}$Ds &-2.99  & 7.34  & -3.77   &  $\alpha$  & $\alpha$     &&     $^{279}$Mt &  2.77 &  3.92       &  &  $\alpha$  &     \\
$^{269}$Hs &-0.34  & 7.41  & 0.99    &  $\alpha$  & $\alpha$     &&     $^{275}$Bh &  1.85 &  5.43       &  &  $\alpha$  &     \\
$^{265}$Sg &2.29   & 6.48  & 1.16    &  $\alpha$  & $\alpha$     &&     $^{271}$Db &  5.15 &  8.31       &  &  $\alpha$  &     \\
$^{261}$Rf &4.66   & 5.81  & 1.83    &  $\alpha$  & $\alpha$     &&     $^{267}$Lr &  10.63 & 10.48      &  & SF         &     \\
$^{257}$No &6.41   & 7.57  & 1.39    &  $\alpha$  & $\alpha$     &&     $^{263}$Md &  11.64 & 11.17      &  & SF         &     \\
\cline{1-6}\cline{8-13}
$^{272}$Ds   &-3.57   & 4.58       &           &$\alpha$   &           && $^{278}$Mt & 3.32&4.64     &0.90 &  $\alpha$   & $\alpha$   \\
$^{268}$Hs   &-2.15   & 3.75       &-0.40     &$\alpha$   & $\alpha$   && $^{274}$Bh & 2.28&7.77     &1.73 &  $\alpha$   & $\alpha$   \\
$^{264}$Sg   &1.63     & 3.28       & -1.43   &$\alpha$   &SF          && $^{270}$Db & 5.46&10.25    &3.56  &  $\alpha$  &  $\alpha$  \\
$^{260}$Rf   &3.55     & 2.86       & -1.68   &SF         & SF         && $^{266}$Lr & 7.48&12.03    &4.60  &  $\alpha$  & SF         \\
$^{256}$No   &4.93     & 6.65       & 0.46    &$\alpha$   & SF         && $^{262}$Md & 9.43&11.04    &      & $\alpha$   &            \\
\cline{1-6}\cline{8-13}
$^{271}$Ds   &-2.33   &5.18   &-2.79    &$\alpha$   &$\alpha$          &&    $^{277}$Mt & 0.72& 4.48    &0.70 &  $\alpha$   & SF  \\
$^{267}$Hs   &-1.58   &5.15   &-1.28    &$\alpha$   &$\alpha$          &&    $^{273}$Bh & 0.62& 8.44    & &  $\alpha$   &         \\
$^{263}$Sg   &1.43    &5.64   & 0.00    &$\alpha$   &$\alpha$          &&    $^{269}$Db & 5.65& 10.95   & &  $\alpha$  &          \\
$^{259}$Rf   &3.43    &5.02   & 0.38    &$\alpha$   &$\alpha$          &&    $^{265}$Lr & 9.51& 10.93   & &  $\alpha$  &          \\
$^{255}$No   &4.62    &9.81   & 2.32    &$\alpha$   &$\alpha$          &&    $^{261}$Md &11.00& 9.89    & & SF         &          \\
\cline{1-6}\cline{8-13}
\end{tabular}}
\label{Ds-chain}
\end{table}

\begin{table}[!htbp]
\caption{Continue to Table \ref{Ds-chain}.}
\centering
\resizebox{1.0\textwidth}{!}{%
\begin{tabular}{|c|c|c|c|c|c|c|c|c|c|c|c|c|}
\cline{1-6}\cline{8-13}
 \multicolumn{1}{|c}{Nucleus}&
 \multicolumn{3}{|c}{log$_{10}$T$_{1/2}$(s)}&
\multicolumn{2}{|c}{Decay Modes}&
 \multicolumn{1}{|c|}{}&
  \multicolumn{1}{|c}{Nucleus}&
 \multicolumn{3}{|c}{log$_{10}$T$_{1/2}$(s)}&
  \multicolumn{2}{|c|}{Decay Modes}\\
  \cline{2-6}\cline{9-13}
 \multicolumn{1}{|c}{}&
 \multicolumn{1}{|c}{$\alpha$(MMF)}& \multicolumn{1}{|c}{SF}&
  \multicolumn{1}{|c}{Expt.}&\multicolumn{1}{|c}{Predicted}&\multicolumn{1}{|c|}{Expt.}&&
\multicolumn{1}{|c}{}&
 \multicolumn{1}{|c}{$\alpha$(MMF)}& \multicolumn{1}{|c}{SF}&
  \multicolumn{1}{|c|}{Expt.}&\multicolumn{1}{c}{Predicted}&\multicolumn{1}{|c|}{Expt.}\\
\cline{1-6}\cline{8-13}
  $^{276}$Mt & 1.43&6.46    &-0.14 &  $\alpha$  &  $\alpha$      &  &   $^{267}$Mt  &-3.16   & 3.55  &         &$\alpha$    & \\
  $^{272}$Bh & 1.74&10.52   &1.00  &  $\alpha$  &  $\alpha$      &  &   $^{263}$Bh   &-1.66  & 5.35  &         &$\alpha$    & \\
  $^{268}$Db & 4.30&12.54   &5.06  &  $\alpha$  &  SF            &  &   $^{259}$Db   &1.66   & 5.39  & -0.29   &$\alpha$    & $\alpha$\\
  $^{264}$Lr & 8.27&10.29   &      &  $\alpha$  &                &  &   $^{255}$Lr   &3.37   & 9.76  & 1.49    &$\alpha$    & $\alpha$\\
  $^{260}$Md & 7.89&9.78    &6.44  & $\alpha$   & SF/$\alpha$    &  &   $^{251}$Md   &3.72   & 13.00 & 2.41    &$\alpha$    & $\alpha$\\
\cline{1-6}\cline{8-13}
  $^{275}$Mt &-1.11&7.05    &-2.01 &  $\alpha$  &  $\alpha$  &     &    $^{266}$Mt  &-2.59   & 3.39   & -2.77  & $\alpha$   & $\alpha$\\
  $^{271}$Bh &0.71 &11.17   &      &  $\alpha$  &            &     &    $^{262}$Bh   &-0.82  & 5.65   & -1.66  & $\alpha$   & $\alpha$\\
  $^{267}$Db &5.29 &11.66   &3.64  &  $\alpha$  &  SF        &     &    $^{258}$Db   &1.33   & 6.47   &  0.62  & $\alpha$   & $\alpha$\\
  $^{263}$Lr &8.91 &8.56    &      &        SF  &             &    &    $^{254}$Lr   &2.62   & 10.32  &  1.26   &$\alpha$    &$\alpha$ \\
  $^{259}$Md &9.76 &7.99    &3.76  &       SF   & SF          &    &    $^{250}$Md   &3.42   & 12.23  &  1.40   &$\alpha$    &$\alpha$ \\
\cline{1-6}\cline{8-13}
  $^{274}$Mt &-0.06&8.98    &-0.36 &  $\alpha$  &  $\alpha$     &     & $^{265}$Mt   &-3.46    &2.64   &          & $\alpha$   & \\
  $^{270}$Bh &0.20 &12.39   &1.78  &  $\alpha$  &  $\alpha$     &     & $^{261}$Bh   &-2.08   &4.29   &-1.93     & $\alpha$   &$\alpha$ \\
  $^{266}$Db &5.68 &11.26   &      &  $\alpha$  &               &     & $^{257}$Db   &0.17    &6.34   & 0.36     & $\alpha$   &$\alpha$ \\
  $^{262}$Lr &6.75 &8.90    &4.16  &  $\alpha$  &   SF          &     & $^{253}$Lr   &2.52    &8.08   & -0.24    & $\alpha$   &$\alpha$ \\
  $^{258}$Md &7.50 &8.74    &6.65  &  $\alpha$  & $\alpha$      &     & $^{249}$Md   &2.51    &9.81   & 1.34     & $\alpha$   &$\alpha$ \\
 \cline{1-6}\cline{8-13}
   $^{273}$Mt &-2.18& 9.13        &&  $\alpha$  &                &     & $^{264}$Mt & -2.66     & 3.05            &   &$\alpha$    &           \\
   $^{269}$Bh &0.53 &11.29        &&  $\alpha$  &                &     &  $^{260}$Bh & -1.41    & 4.17      &-1.46    &$\alpha$    &$\alpha$    \\
   $^{265}$Db &4.97 & 8.97        &&  $\alpha$  &                &     &  $^{256}$Db & -0.05    & 6.09      & 0.20    &$\alpha$    &$\alpha$    \\
   $^{261}$Lr &7.60 & 7.53        &3.37 &  SF  &   SF            &     &  $^{252}$Lr & 2.42& 7.36      & -0.44     &$\alpha$    &$\alpha$    \\
   $^{257}$Md &8.17 & 8.70        &4.30 &  $\alpha$  & $\alpha$  &     &  $^{248}$Md & 2.30& 9.31      & 1.11      &$\alpha$    &$\alpha$    \\
 \cline{1-6}\cline{8-13}
$^{272}$Mt     &-2.12  &10.26        &  &$\alpha$    &           &     &   $^{263}$Mt &-3.74    &1.68      &       &$\alpha$    &          \\
$^{268}$Bh     &2.09   &10.28        &  &$\alpha$    &           &     &   $^{259}$Bh &-2.89    &2.77      &       &$\alpha$    &          \\
$^{264}$Db     &4.18   &8.30         &  & $\alpha$   &           &     &   $^{255}$Db &-0.50    &4.11      &0.20   &$\alpha$    &$\alpha$  \\
$^{260}$Lr     &6.13   &7.82     &2.26  & $\alpha$   &  $\alpha$ &     &   $^{251}$Lr &1.23     &5.19      &       &$\alpha$    &          \\
$^{256}$Md     &6.43   &10.94    &3.66  &$\alpha$    & $\alpha$  &     &   $^{247}$Md &1.92     &7.49      &0.08   &$\alpha$    &$\alpha$  \\
 \cline{1-6}\cline{8-13}
$^{271}$Mt   & -2.28  &8.89    &       &  $\alpha$   &                 &     &  $^{262}$Mt &-3.18    & 1.35       &  &$\alpha$     &         \\
$^{267}$Bh   & -0.34  &8.75    & 1.23   & $\alpha$   & $\alpha$        &     &  $^{258}$Bh &-2.40    & 2.33       &  &$\alpha$     &         \\
$^{263}$Db   & 3.71   &6.90    & 1.43   & $\alpha$   & $\alpha$        &     &  $^{254}$Db &0.24     & 3.13       &  &$\alpha$     &         \\
$^{259}$Lr   & 6.45   &6.36    & 0.79   & SF/$\alpha$   & $\alpha$     &     &  $^{250}$Lr &0.83     & 4.52       &  &$\alpha$     &         \\
$^{255}$Md   & 6.38   &11.56   & 3.21    &$\alpha$   & $\alpha$        &     &  $^{246}$Md &1.32     & 6.82       & -0.05 &$\alpha$&$\alpha$ \\
  \cline{1-6}\cline{8-13}
$^{270}$Mt    & -0.81 &8.11    &-2.30   & $\alpha$   & $\alpha$    &     &    $^{261}$Mt     &-4.39   & -0.42       &  &$\alpha$   &           \\
$^{266}$Bh    & 0.18  &8.21    &0.23    & $\alpha$   & $\alpha$    &     &    $^{257}$Bh     &-3.06   & 0.34        &  &$\alpha$   &           \\
$^{262}$Db    & 3.38  &7.29    &1.54    & $\alpha$   & $\alpha$    &     &    $^{253}$Db     &-1.56   & 0.96        &  &$\alpha$   &           \\
$^{258}$Lr    & 5.23  &7.63    &0.61    & $\alpha$   & $\alpha$    &     &    $^{249}$Lr     &0.44    & 2.29        &  &$\alpha$   &           \\
$^{254}$Md    & 5.04  &13.37   &3.23     &$\alpha$   &             &     &    $^{245}$Md     &0.18    & 4.80   &-3.05  &$\alpha$   &SF/$\alpha$\\
 \cline{1-6}\cline{8-13}
$^{269}$Mt   &-2.73 &6.00    &        & $\alpha$   &             &    &      $^{260}$Mt      & -3.78  & -1.48       &   &$\alpha$   &       \\
$^{265}$Bh   &-0.91 &6.26    &-0.05   & $\alpha$   &$\alpha$     &     &     $^{256}$Bh      & -2.32  & -0.71       &   &$\alpha$   &        \\
$^{261}$Db   &2.57  &5.91    &0.26    & $\alpha$   &$\alpha$     &     &     $^{252}$Db      & -1.12  & 0.13        &   &$\alpha$   &        \\
$^{257}$Lr   &5.07  &8.06    &0.60    & $\alpha$   &$\alpha$     &     &     $^{248}$Lr      & -0.04  & 1.59        &   &$\alpha$   &      \\
$^{253}$Md   &4.53  &13.83   &2.56    & $\alpha$   &$\alpha$     &     &     $^{244}$Md & -0.02  & 4.02        &   &$\alpha$   &       \\
 \cline{1-6}\cline{8-13}
 $^{268}$Mt   &-1.82  &5.03    &-1.68    &$\alpha$  & $\alpha$        &   &  $^{259}$Mt     & -4.48  &  -3.52      &   &$\alpha$   &      \\
     $^{264}$Bh   &-0.30  &6.71    &-0.36    &$\alpha$  & $\alpha$    &   &  $^{255}$Bh     & -3.52  &  -2.76      &   &$\alpha$   &      \\
      $^{260}$Db   &2.37   &6.34   &  0.18    &  $\alpha$  & $\alpha$ &   &  $^{251}$Db     & -2.42  &  -1.94      &   &$\alpha$   &      \\
      $^{256}$Lr   &4.09   &9.51   &  1.43   &  $\alpha$  & $\alpha$  &   &  $^{247}$Lr     & -1.16  & -0.17      &   &$\alpha$   &      \\
      $^{252}$Md   &3.47   &14.75  &  2.14    & $\alpha$  &           &   &  $^{243}$Md     & -0.54  & 2.10       &  &$\alpha$   &      \\
\cline{1-6}\cline{8-13}
\end{tabular}}
\label{Mt-chain}
\end{table}

\section{Conclusions}
Structural properties are explored within a wide range of transfermium isotopes including even and odd nuclei with 101$\leq$Z$\leq$110 and N$\approx$135$-$186. The calculations are performed using relativistic mean-field theory with NL3* parameter and found with a good match with Hartree-Fock-Bogoliubov (HFB) mass model (HFB-24 functional), global mass formula (WS4) and Finite Range Droplet Model (FRDM) calculations along with available experimental data. N$=$184 is found with a spherical shell gap and own to possess similar characteristics as that of other conventional magic numbers. On the other hand, N$=$162 and N$=$152 are found as deformed shell closures which are demonstrated by Nilsson single-particle states. Most of the nuclei in this region are found prolately deformed up to N$\approx$168. The phenomenon of shape transition and shape co-existence is observed for the nuclei with 170$\leq$N$\leq$184.\par

To find the probable decay modes for all these nuclei, first, the $\alpha$-decay half-lives are compared with experimental half-lives using 10 empirical formula of $\alpha$-decay, out of which the formula of Manjunatha \textit{et al.} is found with the best match. This formula is further modified in this article by adding terms of asymmetry ($I$ and $I^2$) and the new coefficients are obtained by using 366 experimental $\alpha$-decay half-lives. This modified formula has reproduced the experimental data quite well, in fact, better than the original formula of Manjunatha \textit{et al.}. Therefore, this modified Manjunatha formula (MMF-2020) is utilized to examine the competition between $\alpha$-decay and spontaneous fission and consequently to construct $\alpha$-decay chains for this region of the periodic chart. A total of 18 decay chains for Ds and 22 decay chains of Mt are constituted with their probable decay modes and $\alpha$-decay half-lives. The predicted decay modes and half-lives are found in excellent agreement with available experimental data. Hence, this study may provide useful inputs for future experiments and search of new elements in this transfermium region.\par

\section{Acknowledgement}
The authors take great pleasure in thanking the referee for his several suggestions and comments which helped to improve the manuscript. G. Saxena acknowledges the support provided by SERB (DST), Govt. of India under CRG/2019/001851.

\section{Appendix}
In this section, few ground state properties viz. binding energy (B.E.), charge radius $R_c$, neutron radius $R_n$, proton radius $R_p$, matter radius $R_m$, quadrupole deformation parameter $\beta$ are given in the following tables for transfermium isotopes for the range proton number 101$\leq$Z$\leq$110 and neutron number N$\approx$135$-$186. Binding energies shown are in the unit of MeV and all the radii are in the unit of fm. Positive and negative signs in the value of $\beta$ represent prolate and oblate deformation, respectively.

\begin{table}[!htbp]
\caption{Ground state properties viz. binding energy (B.E.), charge radius $R_c$, neutron radius $R_n$, proton radius $R_p$, matter radius $R_m$, quadrupole deformation parameter $\beta$ for Md and No isotopes.}
\centering
\resizebox{1.0\textwidth}{!}{%
{\begin{tabular}{ccccccccccccccccc}

\cline{1-8}\cline{10-17}
 \multicolumn{1}{c}{Nucleus}&
 \multicolumn{2}{c}{B.E.}&
  \multicolumn{1}{c}{}&\multicolumn{1}{c}{}& \multicolumn{1}{c}{}& \multicolumn{1}{c}{}& \multicolumn{1}{c}{}&&
  \multicolumn{1}{c}{Nucleus}&
 \multicolumn{2}{c}{B.E.}&
  \multicolumn{1}{c}{}&\multicolumn{1}{c}{}& \multicolumn{1}{c}{}& \multicolumn{1}{c}{}& \multicolumn{1}{c}{}\\
   \cline{2-3} \cline{11-12}
 \multicolumn{1}{c}{}&
 \multicolumn{1}{c}{RMF}& \multicolumn{1}{c}{Expt.}&
  \multicolumn{1}{c}{$R_c$}&\multicolumn{1}{c}{$R_n$}&\multicolumn{1}{c}{$R_p$}& \multicolumn{1}{c}{$R_m$}& \multicolumn{1}{c}{$\beta$}&&
  \multicolumn{1}{c}{ }&
 \multicolumn{1}{c}{RMF}& \multicolumn{1}{c}{Expt.}&
  \multicolumn{1}{c}{$R_c$}&\multicolumn{1}{c}{$R_n$}&\multicolumn{1}{c}{$R_p$}& \multicolumn{1}{c}{$R_m$}& \multicolumn{1}{c}{$\beta$}\\
\cline{1-8}\cline{10-17}
$^{235}$Md & 1746.455 & & 5.942 & 6.004 & 5.888 & 5.954 & 0.318&& $^{238}$No & 1765.421 &  & 5.952 & 6.016 & 5.898 & 5.966& 0.294 \\
$^{236}$Md & 1754.351 & & 5.917 & 5.983 & 5.863 & 5.932 & 0.258&& $^{239}$No & 1773.372 &  & 5.939 & 6.008 & 5.885 & 5.956& 0.251 \\
$^{237}$Md & 1764.255 & & 5.936 & 6.009 & 5.882 & 5.955 & 0.287&& $^{240}$No & 1783.179 &  & 5.946 & 6.023 & 5.892 & 5.967& 0.254 \\
$^{238}$Md & 1772.017 & & 5.925 & 6.005 & 5.871 & 5.948 & 0.252&& $^{241}$No & 1790.625 &  & 5.964 & 6.046 & 5.910 & 5.989& 0.283 \\
$^{239}$Md & 1781.604 & & 5.932 & 6.019 & 5.878 & 5.960 & 0.254&& $^{242}$No & 1800.094 &  & 5.957 & 6.047 & 5.903 & 5.987& 0.257 \\
$^{240}$Md & 1788.868 & & 5.949 & 6.041 & 5.895 & 5.980 & 0.281&& $^{243}$No & 1807.317 &  & 5.962 & 6.057 & 5.908 & 5.995& 0.256 \\
$^{241}$Md & 1798.111 & & 5.955 & 6.055 & 5.901 & 5.991 & 0.282&& $^{244}$No & 1816.540 &  & 5.970 & 6.073 & 5.916 & 6.008& 0.263 \\
$^{242}$Md & 1805.178 & & 5.949 & 6.054 & 5.895 & 5.988 & 0.258&& $^{245}$No & 1823.370 &  & 5.975 & 6.083 & 5.921 & 6.016& 0.266 \\
$^{243}$Md & 1814.179 & & 5.957 & 6.069 & 5.903 & 6.001 & 0.264&& $^{246}$No & 1832.406 &  & 5.983 & 6.098 & 5.929 & 6.028& 0.270 \\
$^{244}$Md & 1820.769 & & 5.962 & 6.080 & 5.908 & 6.009 & 0.266&& $^{247}$No & 1839.102 &  & 5.988 & 6.109 & 5.935 & 6.038& 0.272 \\
$^{245}$Md & 1829.554 & 1823.184& 5.970 & 6.094 & 5.916 & 6.022 & 0.269&& $^{248}$No & 1847.853 & 1841.266 & 6.003 & 6.131 & 5.950 & 6.057& 0.290 \\
$^{246}$Md & 1836.038 & 1830.412& 5.976 & 6.105 & 5.922 & 6.031 & 0.272&& $^{249}$No & 1854.196 & 1848.177 & 6.000 & 6.132 & 5.947 & 6.057& 0.275 \\
$^{247}$Md & 1844.545 & 1838.661& 5.983 & 6.120 & 5.929 & 6.043 & 0.275&& $^{250}$No & 1862.538 & 1856.466 & 6.008 & 6.148 & 5.954 & 6.070& 0.280 \\
$^{248}$Md & 1850.703 & 1845.522& 5.988 & 6.129 & 5.934 & 6.050 & 0.275&& $^{251}$No & 1868.801 & 1863.252 & 6.019 & 6.165 & 5.966 & 6.085& 0.293 \\
$^{249}$Md & 1858.746 & 1853.509& 6.002 & 6.151 & 5.948 & 6.070 & 0.289&& $^{252}$No & 1876.570 & 1871.301 & 6.026 & 6.179 & 5.973 & 6.096& 0.292 \\
$^{250}$Md & 1864.705 & 1860.182& 6.007 & 6.162 & 5.953 & 6.079 & 0.290&& $^{253}$No & 1882.276 & 1877.885 & 6.031 & 6.189 & 5.978 & 6.105& 0.292 \\
$^{251}$Md & 1872.239 & 1867.917& 6.014 & 6.176 & 5.960 & 6.090 & 0.289&& $^{254}$No & 1889.775 & 1885.592 & 6.038 & 6.203 & 5.985 & 6.116& 0.290 \\
$^{252}$Md & 1877.659 & 1874.445& 6.019 & 6.186 & 5.965 & 6.098 & 0.289&& $^{255}$No & 1895.267 & 1891.579 & 6.043 & 6.212 & 5.99 & 6.124& 0.288 \\
$^{253}$Md & 1884.920 & 1881.853& 6.020 & 6.193 & 5.967 & 6.104 & 0.276&& $^{256}$No & 1902.437 & 1898.636 & 6.05 & 6.227 & 5.997 & 6.137& 0.287 \\
$^{254}$Md & 1890.143 & 1887.645& 6.031 & 6.210 & 5.978 & 6.119 & 0.286&& $^{257}$No & 1907.683 & 1904.282 & 6.056 & 6.238 & 6.003 & 6.146& 0.286 \\
$^{255}$Md & 1897.099 & 1894.326& 6.032 & 6.218 & 5.979 & 6.124 & 0.273&& $^{258}$No & 1914.519 & 1911.122 & 6.056 & 6.246 & 6.003 & 6.151& 0.274 \\
$^{256}$Md & 1902.089 & 1899.784& 6.038 & 6.229 & 5.985 & 6.134 & 0.273&& $^{259}$No & 1919.521 & 1916.593 & 6.062 & 6.256 & 6.009 & 6.160& 0.273 \\
$^{257}$Md & 1908.715 & 1906.318& 6.044 & 6.243 & 5.991 & 6.146 & 0.271&& $^{260}$No & 1926.029 & 1923.131 & 6.068 & 6.270 & 6.015 & 6.172& 0.271 \\
$^{258}$Md & 1913.441 & 1911.696& 6.050 & 6.254 & 5.997 & 6.155 & 0.270&& $^{261}$No & 1930.628 & 1928.359 & 6.073 & 6.280 & 6.020 & 6.180& 0.270 \\
$^{259}$Md & 1919.734 & 1917.830& 6.056 & 6.268 & 6.003 & 6.166 & 0.268&& $^{262}$No & 1936.934 & 1934.785 & 6.080 & 6.294 & 6.027 & 6.191& 0.267 \\
$^{260}$Md & 1924.054 & 1922.973& 6.061 & 6.278 & 6.008 & 6.175 & 0.267&& $^{263}$No & 1941.427 & 1939.828 & 6.085 & 6.303 & 6.032 & 6.199& 0.265 \\
$^{261}$Md & 1930.225 & 1929.019& 6.068 & 6.292 & 6.015 & 6.186 & 0.264&& $^{264}$No & 1947.221 & 1946.017 & 6.091 & 6.317 & 6.038 & 6.211& 0.261 \\
$^{262}$Md & 1934.405 & 1934.041& 6.073 & 6.302 & 6.020 & 6.195 & 0.262&& $^{265}$No & 1951.123 &  & 6.095 & 6.324 & 6.042 & 6.217& 0.250 \\
$^{263}$Md & 1940.111 & & 6.079 & 6.316 & 6.027 & 6.206 & 0.259&& $^{266}$No & 1957.031 &  & 6.104 & 6.343 & 6.051 & 6.233& 0.256 \\
$^{264}$Md & 1943.904 & & 6.084 & 6.323 & 6.031 & 6.213 & 0.250&& $^{267}$No & 1960.590 &  & 6.108 & 6.352 & 6.055 & 6.240& 0.249 \\
$^{265}$Md & 1949.683 & & 6.093 & 6.342 & 6.040 & 6.228 & 0.254&& $^{268}$No & 1966.142 &  & 6.117 & 6.368 & 6.064 & 6.254& 0.250 \\
$^{266}$Md & 1953.103 & & 6.097 & 6.351 & 6.044 & 6.236 & 0.247&& $^{269}$No & 1969.393 &  & 6.101 & 6.354 & 6.049 & 6.240& 0.191 \\
$^{267}$Md & 1958.590 & & 6.106 & 6.367 & 6.053 & 6.250 & 0.250&& $^{270}$No & 1974.931 &  & 6.129 & 6.393 & 6.076 & 6.275& 0.245 \\
$^{268}$Md & 1961.659 & & 6.090 & 6.353 & 6.037 & 6.236 & 0.189&& $^{271}$No & 1978.447 &  & 6.118 & 6.385 & 6.065 & 6.266& 0.200 \\
$^{269}$Md & 1967.150 & & 6.104 & 6.376 & 6.051 & 6.256 & 0.210&& $^{272}$No & 1983.853 &  & 6.114 & 6.388 & 6.061 & 6.267& 0.163 \\
$^{270}$Md & 1970.517 & & 6.107 & 6.385 & 6.054 & 6.263 & 0.199&& $^{273}$No & 1987.309 &  & 6.117 & 6.396 & 6.064 & 6.274& 0.154 \\
$^{271}$Md & 1973.414 & & 6.090 & 6.370 & 6.038 & 6.249 & -0.109&&$^{274}$No & 1993.045 &  & 6.127 & 6.413 & 6.074 & 6.289& -0.157 \\
$^{272}$Md & 1979.290 & & 6.110& 6.399 & 6.057 & 6.274 & -0.159&& $^{275}$No & 1996.729 &  & 6.125 & 6.415 & 6.073 & 6.290& -0.130 \\
$^{273}$Md & 1985.013 & & 6.110 & 6.407 & 6.057 & 6.280 & -0.138&& $^{276}$No & 2002.136 &  & 6.130 & 6.428 & 6.077 & 6.300& -0.124 \\
$^{274}$Md & 1988.584 & & 6.119 & 6.422 & 6.066 & 6.293 & -0.153&& $^{277}$No & 2005.645 &  & 6.133 & 6.436 & 6.080 & 6.307& -0.120 \\
$^{275}$Md & 1993.781 & & 6.124 & 6.435 & 6.072 & 6.304 & -0.150&& $^{278}$No & 2010.788 &  & 6.138 & 6.449 & 6.085 & 6.318& -0.115 \\
$^{276}$Md & 1997.176 & & 6.122 & 6.437 & 6.069 & 6.305 & -0.124&& $^{279}$No & 2014.146 &  & 6.140 & 6.457 & 6.088 & 6.324& -0.111 \\
$^{277}$Md & 2002.183 & & 6.126 & 6.451 & 6.074 & 6.316 & -0.119&& $^{280}$No & 2018.990 &  & 6.145 & 6.471 & 6.093 & 6.336& -0.106 \\
$^{278}$Md & 2005.335 & & 6.130 & 6.457 & 6.077 & 6.322 & -0.111&& $^{281}$No & 2022.154 &  & 6.143 & 6.477 & 6.091 & 6.339& -0.071 \\
$^{279}$Md & 2010.054 & & 6.134 & 6.472 & 6.082 & 6.333 & -0.105&& $^{282}$No & 2026.665 &  & 6.151 & 6.494 & 6.099 & 6.354& -0.085 \\
$^{280}$Md & 2013.099 & & 6.132 & 6.478 & 6.080 & 6.337 & -0.072&& $^{283}$No & 2029.902 &  & 6.149 & 6.500 & 6.096 & 6.358& -0.070 \\
$^{281}$Md & 2017.494 & & 6.140 & 6.495 & 6.088 & 6.351 & -0.085&& $^{284}$No & 2034.270 &  & 6.150 & 6.513 & 6.098 & 6.367& 0.025 \\
$^{282}$Md & 2020.623 & & 6.138 & 6.501 & 6.085 & 6.356 & -0.070&& $^{285}$No & 2037.639 &  & 6.151 & 6.523 & 6.099 & 6.374& -0.029 \\
$^{283}$Md & 2024.970 & & 6.140 & 6.514 & 6.088 & 6.365 & -0.041&& $^{286}$No & 2041.710 &  & 6.153 & 6.536 & 6.101 & 6.384& 0.000 \\
$^{284}$Md & 2028.235 & & 6.140 & 6.524 & 6.088 & 6.373 & -0.031&& $^{287}$No & 2043.459 &  & 6.165 & 6.549 & 6.113 & 6.398& -0.018 \\
$^{285}$Md & 2032.110 & & 6.143 & 6.537 & 6.090 & 6.383 & 0.000&& $^{288}$No & 2047.112 &  & 6.179 & 6.566 & 6.127 & 6.414& 0.025 \\
$^{286}$Md & 2033.705 & & 6.156 & 6.552 & 6.103 & 6.397 & -0.030&&   &  &  &  &  &  & &  \\
$^{287}$Md & 2037.146 & & 6.168 & 6.568 & 6.116 & 6.413 & 0.025&&  &  &  &  &  &  & &  \\
\cline{1-8}\cline{10-17}
\end{tabular}}}
\label{gs-properties-Md-Lo}
\end{table}

\begin{table}[!htbp]
\caption{Same as Table \ref{gs-properties-Md-Lo} but for Lr and Rf isotopes.}
\centering
\resizebox{1.0\textwidth}{!}{%
{\begin{tabular}{ccccccccccccccccc}

\cline{1-8}\cline{10-17}
 \multicolumn{1}{c}{Nucleus}&
 \multicolumn{2}{c}{B.E.}&
  \multicolumn{1}{c}{}&\multicolumn{1}{c}{}& \multicolumn{1}{c}{}& \multicolumn{1}{c}{}& \multicolumn{1}{c}{}&&
  \multicolumn{1}{c}{Nucleus}&
 \multicolumn{2}{c}{B.E.}&
  \multicolumn{1}{c}{}&\multicolumn{1}{c}{}& \multicolumn{1}{c}{}& \multicolumn{1}{c}{}& \multicolumn{1}{c}{}\\
   \cline{2-3} \cline{11-12}
 \multicolumn{1}{c}{}&
 \multicolumn{1}{c}{RMF}& \multicolumn{1}{c}{Expt.}&
  \multicolumn{1}{c}{$R_c$}&\multicolumn{1}{c}{$R_n$}&\multicolumn{1}{c}{$R_p$}& \multicolumn{1}{c}{$R_m$}& \multicolumn{1}{c}{$\beta$}&&
  \multicolumn{1}{c}{ }&
 \multicolumn{1}{c}{RMF}& \multicolumn{1}{c}{Expt.}&
  \multicolumn{1}{c}{$R_c$}&\multicolumn{1}{c}{$R_n$}&\multicolumn{1}{c}{$R_p$}& \multicolumn{1}{c}{$R_m$}& \multicolumn{1}{c}{$\beta$}\\
\cline{1-8}\cline{10-17}

$^{	241	}$Lr	&	1782.794	&		&	5.957	&	6.025	&	5.903	&	5.973	&	0.254	&&	$^{	242	}$Rf	&	1783.612	&		&	5.970	&	 6.029	&	5.916	&	5.981	&	0.253	\\
$^{	242	}$Lr	&	1790.447	&		&	5.971	&	6.045	&	5.917	&	5.991	&	0.272	&&	$^{	243	}$Rf	&	1791.314	&		&	5.994	&	 6.058	&	5.940	&	6.008	&	0.293	\\
$^{	243	}$Lr	&	1800.143	&		&	5.984	&	6.066	&	5.930	&	6.009	&	0.297	&&	$^{	244	}$Rf	&	1801.330	&		&	6.000	&	 6.072	&	5.947	&	6.019	&	0.296	\\
$^{	244	}$Lr	&	1807.572	&		&	5.972	&	6.059	&	5.919	&	6.000	&	0.256	&&	$^{	245	}$Rf	&	1809.009	&		&	6.002	&	 6.080	&	5.949	&	6.025	&	0.291	\\
$^{	245	}$Lr	&	1817.142	&		&	5.993	&	6.089	&	5.940	&	6.026	&	0.295	&&	$^{	246	}$Rf	&	1818.676	&		&	5.994	&	 6.078	&	5.940	&	6.020	&	0.261	\\
$^{	246	}$Lr	&	1824.101	&		&	6.006	&	6.108	&	5.953	&	6.043	&	0.309	&&	$^{	247	}$Rf	&	1825.865	&		&	5.998	&	 6.089	&	5.945	&	6.028	&	0.264	\\
$^{	247	}$Lr	&	1833.576	&		&	6.003	&	6.111	&	5.950	&	6.044	&	0.294	&&	$^{	248	}$Rf	&	1835.476	&		&	6.019	&	 6.116	&	5.965	&	6.053	&	0.293	\\
$^{	248	}$Lr	&	1840.393	&		&	6.014	&	6.127	&	5.961	&	6.058	&	0.303	&&	$^{	249	}$Rf	&	1842.495	&		&	6.022	&	 6.124	&	5.968	&	6.059	&	0.289	\\
$^{	249	}$Lr	&	1849.501	&		&	6.014	&	6.134	&	5.960	&	6.062	&	0.294	&&	$^{	250	}$Rf	&	1851.762	&		&	6.029	&	 6.138	&	5.975	&	6.071	&	0.293	\\
$^{	250	}$Lr	&	1855.979	&		&	6.011	&	6.134	&	5.957	&	6.062	&	0.279	&&	$^{	251	}$Rf	&	1858.506	&		&	6.023	&	 6.136	&	5.970	&	6.068	&	0.273	\\
$^{	251	}$Lr	&	1864.723	&	1857.591	&	6.024	&	6.156	&	5.971	&	6.081	&	0.294	&&	$^{	252	}$Rf	&	1867.387	&		&	6.039	 &	6.160	&	5.986	&	6.089	&	0.294	\\
$^{	252	}$Lr	&	1871.257	&	1864.653	&	6.029	&	6.166	&	5.976	&	6.089	&	0.294	&&	$^{	253	}$Rf	&	1874.162	&	1867.122	&	 6.043	&	6.170	&	5.990	&	6.097	&	0.295	\\
$^{	253	}$Lr	&	1879.301	&	1872.887	&	6.036	&	6.180	&	5.983	&	6.100	&	0.293	&&	$^{	254	}$Rf	&	1882.464	&	1875.552	&	 6.050	&	6.184	&	5.997	&	6.108	&	0.294	\\
$^{	254	}$Lr	&	1885.262	&	1879.662	&	6.040	&	6.189	&	5.987	&	6.108	&	0.293	&&	$^{	255	}$Rf	&	1888.653	&	1882.492	&	 6.054	&	6.193	&	6.000	&	6.115	&	0.294	\\
$^{	255	}$Lr	&	1893.017	&	1887.657	&	6.047	&	6.203	&	5.994	&	6.120	&	0.292	&&	$^{	256	}$Rf	&	1896.666	&	1890.671	&	 6.061	&	6.207	&	6.008	&	6.127	&	0.293	\\
$^{	256	}$Lr	&	1898.748	&	1893.929	&	6.053	&	6.213	&	5.999	&	6.128	&	0.289	&&	$^{	257	}$Rf	&	1902.633	&	1897.098	&	 6.066	&	6.216	&	6.013	&	6.134	&	0.290	\\
$^{	257	}$Lr	&	1906.139	&	1901.082	&	6.060	&	6.228	&	6.007	&	6.140	&	0.289	&&	$^{	258	}$Rf	&	1910.272	&	1904.695	&	 6.073	&	6.231	&	6.020	&	6.147	&	0.290	\\
$^{	258	}$Lr	&	1911.610	&	1907.036	&	6.065	&	6.238	&	6.012	&	6.149	&	0.287	&&	$^{	259	}$Rf	&	1915.971	&	1910.745	&	 6.078	&	6.241	&	6.025	&	6.155	&	0.288	\\
$^{	259	}$Lr	&	1918.641	&	1914.037	&	6.072	&	6.253	&	6.019	&	6.161	&	0.286	&&	$^{	260	}$Rf	&	1923.217	&	1918.031	&	 6.085	&	6.255	&	6.032	&	6.167	&	0.287	\\
$^{	260	}$Lr	&	1923.828	&	1919.684	&	6.077	&	6.263	&	6.024	&	6.169	&	0.284	&&	$^{	261	}$Rf	&	1928.686	&	1923.931	&	 6.085	&	6.259	&	6.032	&	6.170	&	0.276	\\
$^{	261	}$Lr	&	1930.503	&	1926.474	&	6.078	&	6.271	&	6.025	&	6.175	&	0.274	&&	$^{	262	}$Rf	&	1935.589	&	1930.928	&	 6.091	&	6.273	&	6.038	&	6.181	&	0.274	\\
$^{	262	}$Lr	&	1935.334	&	1932.001	&	6.088	&	6.286	&	6.035	&	6.189	&	0.282	&&	$^{	263	}$Rf	&	1940.697	&	1936.636	&	 6.102	&	6.289	&	6.049	&	6.195	&	0.282	\\
$^{	263	}$Lr	&	1941.789	&	1938.446	&	6.089	&	6.294	&	6.037	&	6.194	&	0.270	&&	$^{	264	}$Rf	&	1947.358	&	1943.387	&	 6.102	&	6.296	&	6.050	&	6.200	&	0.270	\\
$^{	264	}$Lr	&	1946.465	&	1943.869	&	6.094	&	6.303	&	6.041	&	6.202	&	0.268	&&	$^{	265	}$Rf	&	1952.332	&	1948.845	&	 6.107	&	6.304	&	6.054	&	6.207	&	0.268	\\
$^{	265	}$Lr	&	1952.386	&	1950.084	&	6.101	&	6.317	&	6.048	&	6.214	&	0.265	&&	$^{	266	}$Rf	&	1958.372	&	1955.530	&	 6.114	&	6.318	&	6.061	&	6.219	&	0.265	\\
$^{	266	}$Lr	&	1956.375	&	1954.767	&	6.105	&	6.324	&	6.052	&	6.220	&	0.255	&&	$^{	267	}$Rf	&	1962.502	&	1960.234	&	 6.117	&	6.325	&	6.064	&	6.225	&	0.253	\\
$^{	267	}$Lr	&	1962.477	&		&	6.115	&	6.344	&	6.062	&	6.237	&	0.262	&&	$^{	268	}$Rf	&	1968.770	&	1966.273	&	6.127	 &	6.345	&	6.075	&	6.241	&	0.260	\\
$^{	268	}$Lr	&	1966.143	&		&	6.119	&	6.353	&	6.067	&	6.245	&	0.256	&&	$^{	269	}$Rf	&	1972.666	&		&	6.131	&	 6.354	&	6.078	&	6.249	&	0.253	\\
$^{	269	}$Lr	&	1971.848	&		&	6.128	&	6.370	&	6.076	&	6.259	&	0.259	&&	$^{	270	}$Rf	&	1978.449	&		&	6.140	&	 6.370	&	6.088	&	6.263	&	0.256	\\
$^{	270	}$Lr	&	1975.280	&		&	6.147	&	6.396	&	6.094	&	6.283	&	0.282	&&	$^{	271	}$Rf	&	1981.986	&		&	6.131	&	 6.362	&	6.078	&	6.255	&	0.215	\\
$^{	271	}$Lr	&	1980.900	&		&	6.142	&	6.396	&	6.090	&	6.281	&	0.256	&&	$^{	272	}$Rf	&	1987.850	&		&	6.137	&	 6.377	&	6.085	&	6.267	&	0.212	\\
$^{	272	}$Lr	&	1984.376	&		&	6.127	&	6.384	&	6.075	&	6.268	&	0.199	&&	$^{	273	}$Rf	&	1991.583	&		&	6.139	&	 6.384	&	6.087	&	6.273	&	0.201	\\
$^{	273	}$Lr	&	1989.755	&		&	6.134	&	6.398	&	6.081	&	6.280	&	0.198	&&	$^{	274	}$Rf	&	1997.162	&		&	6.145	&	 6.398	&	6.093	&	6.284	&	0.197	\\
$^{	274	}$Lr	&	1993.246	&		&	6.128	&	6.395	&	6.075	&	6.277	&	0.156	&&	$^{	275	}$Rf	&	2000.816	&		&	6.146	&	 6.405	&	6.094	&	6.289	&	0.184	\\
$^{	275	}$Lr	&	1998.877	&		&	6.138	&	6.412	&	6.085	&	6.292	&	-0.160	&&	$^{	276	}$Rf	&	2006.261	&		&	6.149	&	 6.412	&	6.096	&	6.295	&	-0.159	\\
$^{	276	}$Lr	&	2002.679	&		&	6.141	&	6.421	&	6.089	&	6.299	&	-0.155	&&	$^{	277	}$Rf	&	2010.067	&		&	6.146	&	 6.413	&	6.093	&	6.295	&	-0.131	\\
$^{	277	}$Lr	&	2008.183	&		&	6.147	&	6.434	&	6.095	&	6.310	&	-0.152	&&	$^{	278	}$Rf	&	2015.869	&		&	6.158	&	 6.434	&	6.106	&	6.313	&	-0.151	\\
$^{	278	}$Lr	&	2011.779	&		&	6.150	&	6.442	&	6.097	&	6.317	&	-0.146	&&	$^{	279	}$Rf	&	2019.606	&		&	6.160	&	 6.442	&	6.108	&	6.319	&	-0.144	\\
$^{	279	}$Lr	&	2017.049	&		&	6.148	&	6.448	&	6.096	&	6.321	&	-0.119	&&	$^{	280	}$Rf	&	2024.991	&		&	6.167	&	 6.455	&	6.115	&	6.331	&	-0.142	\\
$^{	280	}$Lr	&	2020.533	&		&	6.151	&	6.456	&	6.099	&	6.327	&	-0.116	&&	$^{	281	}$Rf	&	2028.594	&		&	6.170	&	 6.463	&	6.118	&	6.338	&	-0.138	\\
$^{	281	}$Lr	&	2025.487	&		&	6.157	&	6.470	&	6.105	&	6.339	&	-0.112	&&	$^{	282	}$Rf	&	2033.663	&		&	6.167	&	 6.469	&	6.115	&	6.341	&	-0.110	\\
$^{	282	}$Lr	&	2028.645	&		&	6.167	&	6.485	&	6.115	&	6.352	&	-0.132	&&	$^{	283	}$Rf	&	2037.086	&		&	6.163	&	 6.474	&	6.111	&	6.343	&	-0.072	\\
$^{	283	}$Lr	&	2033.283	&		&	6.157	&	6.488	&	6.105	&	6.352	&	-0.061	&&	$^{	284	}$Rf	&	2041.890	&		&	6.167	&	 6.487	&	6.115	&	6.354	&	-0.060	\\
$^{	284	}$Lr	&	2036.614	&		&	6.157	&	6.496	&	6.105	&	6.357	&	-0.058	&&	$^{	285	}$Rf	&	2045.325	&		&	6.167	&	 6.495	&	6.115	&	6.359	&	-0.057	\\
$^{	285	}$Lr	&	2041.064	&		&	6.161	&	6.511	&	6.109	&	6.369	&	0.027	&&	$^{	286	}$Rf	&	2049.915	&		&	6.171	&	 6.510	&	6.119	&	6.371	&	0.027	\\
$^{	286	}$Lr	&	2044.471	&		&	6.161	&	6.520	&	6.109	&	6.375	&	0.021	&&	$^{	287	}$Rf	&	2053.432	&		&	6.171	&          6.519	&	6.119	&	6.377	&	0.020	\\
$^{	287	}$Lr	&	2048.669	&		&	6.163	&	6.533	&	6.111	&	6.385	&	0.000	&&	$^{	288	}$Rf	&	2057.726	&		&	6.173	&	 6.532	&	6.121	&	6.386	&	0.000	\\
$^{	288	}$Lr	&	2050.536	&		&	6.175	&	6.548	&	6.123	&	6.399	&	-0.030	&&	$^{	289	}$Rf	&	2059.636	&		&	6.185	&	 6.545	&	6.133	&	6.400	&	-0.005	\\
$^{	289	}$Lr	&	2054.470	&		&	6.189	&	6.564	&	6.137	&	6.415	&	0.028	&&	$^{	290	}$Rf	&	2063.778	&		&	6.199	&	 6.563	&	6.148	&	6.417	&	0.027	\\
\cline{1-8}\cline{10-17}
\end{tabular}}
}
\end{table}

\begin{table}[!htbp]
\caption{Same as Table \ref{gs-properties-Md-Lo} but for Db and Sg isotopes.}
\centering
\resizebox{1.0\textwidth}{!}{%
{\begin{tabular}{ccccccccccccccccc}

\cline{1-8}\cline{10-17}
 \multicolumn{1}{c}{Nucleus}&
 \multicolumn{2}{c}{B.E.}&
  \multicolumn{1}{c}{}&\multicolumn{1}{c}{}& \multicolumn{1}{c}{}& \multicolumn{1}{c}{}& \multicolumn{1}{c}{}&&
  \multicolumn{1}{c}{Nucleus}&
 \multicolumn{2}{c}{B.E.}&
  \multicolumn{1}{c}{}&\multicolumn{1}{c}{}& \multicolumn{1}{c}{}& \multicolumn{1}{c}{}& \multicolumn{1}{c}{}\\
   \cline{2-3} \cline{11-12}
 \multicolumn{1}{c}{}&
 \multicolumn{1}{c}{RMF}& \multicolumn{1}{c}{Expt.}&
  \multicolumn{1}{c}{$R_c$}&\multicolumn{1}{c}{$R_n$}&\multicolumn{1}{c}{$R_p$}& \multicolumn{1}{c}{$R_m$}& \multicolumn{1}{c}{$\beta$}&&
  \multicolumn{1}{c}{ }&
 \multicolumn{1}{c}{RMF}& \multicolumn{1}{c}{Expt.}&
  \multicolumn{1}{c}{$R_c$}&\multicolumn{1}{c}{$R_n$}&\multicolumn{1}{c}{$R_p$}& \multicolumn{1}{c}{$R_m$}& \multicolumn{1}{c}{$\beta$}\\
\cline{1-8}\cline{10-17}
$^{245}$Db & 1800.661 & & 6.036 & 6.100 & 5.983 & 6.050 & 0.339&& $^{246}$Sg & 1801.169 &  & 6.056 & 6.109 & 6.003 & 6.064& 0.347 \\
$^{246}$Db & 1808.453 & & 5.995 & 6.064 & 5.941 & 6.012 & 0.250&& $^{247}$Sg & 1808.975 &  & 6.010 & 6.069 & 5.957 & 6.021& 0.254 \\
$^{247}$Db & 1818.373 & & 6.002 & 6.078 & 5.948 & 6.023 & 0.254&& $^{248}$Sg & 1819.126 &  & 6.039 & 6.104 & 5.985 & 6.054& 0.300 \\
$^{248}$Db & 1825.684 & & 6.019 & 6.102 & 5.966 & 6.044 & 0.283&& $^{249}$Sg & 1826.881 &  & 6.042 & 6.112 & 5.988 & 6.060& 0.296 \\
$^{249}$Db & 1835.472 & & 6.025 & 6.114 & 5.971 & 6.054 & 0.282&& $^{250}$Sg & 1836.888 &  & 6.046 & 6.124 & 5.993 & 6.069& 0.295 \\
$^{250}$Db & 1842.843 & & 6.021 & 6.115 & 5.968 & 6.054 & 0.268&& $^{251}$Sg & 1844.179 &  & 6.047 & 6.130 & 5.994 & 6.073& 0.288 \\
$^{251}$Db & 1852.330 & & 6.028 & 6.129 & 5.975 & 6.065 & 0.270&& $^{252}$Sg & 1853.877 &  & 6.054 & 6.145 & 6.001 & 6.085& 0.292 \\
$^{252}$Db & 1859.402 & & 6.054 & 6.160 & 6.001 & 6.095 & 0.312&& $^{253}$Sg & 1861.099 &  & 6.045 & 6.141 & 5.992 & 6.079& 0.267 \\
$^{253}$Db & 1868.570 & & 6.050 & 6.163 & 5.997 & 6.095 & 0.296&& $^{254}$Sg & 1870.374 &  & 6.053 & 6.156 & 6.000 & 6.092& 0.273 \\
$^{254}$Db & 1875.553 & & 6.054 & 6.172 & 6.001 & 6.102 & 0.296&& $^{255}$Sg & 1877.548 &  & 6.068 & 6.176 & 6.015 & 6.109& 0.291 \\
$^{255}$Db & 1884.166 & 1876.446& 6.061 & 6.185 & 6.008 & 6.113 & 0.295&& $^{256}$Sg & 1886.276 &  & 6.074 & 6.188 & 6.021 & 6.119& 0.291 \\
$^{256}$Db & 1890.614 & 1883.613& 6.064 & 6.194 & 6.011 & 6.119 & 0.294&& $^{257}$Sg & 1892.901 &  & 6.077 & 6.196 & 6.025 & 6.126& 0.290 \\
$^{257}$Db & 1898.957 & 1891.975& 6.072 & 6.208 & 6.019 & 6.131 & 0.293&& $^{258}$Sg & 1901.470 & 1865.591 & 6.085 & 6.210 & 6.032 & 6.137& 0.289 \\
$^{258}$Db & 1905.166 & 1898.456& 6.077 & 6.216 & 6.024 & 6.139 & 0.290&& $^{259}$Sg & 1907.892 & 1901.022 & 6.089 & 6.218 & 6.036 & 6.144& 0.286 \\
$^{259}$Db & 1913.154 & 1906.334& 6.084 & 6.232 & 6.031 & 6.152 & 0.290&& $^{260}$Sg & 1916.039 & 1909.066 & 6.096 & 6.233 & 6.044 & 6.156& 0.287 \\
$^{260}$Db & 1919.024 & 1912.723& 6.089 & 6.242 & 6.036 & 6.160 & 0.288&& $^{261}$Sg & 1922.211 & 1915.68 & 6.101 & 6.243 & 6.048 & 6.164& 0.285 \\
$^{261}$Db & 1926.538 & 1920.159& 6.096 & 6.256 & 6.043 & 6.171 & 0.288&& $^{262}$Sg & 1929.977 & 1923.391 & 6.108 & 6.257 & 6.056 & 6.176& 0.285 \\
$^{262}$Db & 1932.233 & 1926.286& 6.101 & 6.266 & 6.049 & 6.180 & 0.286&& $^{263}$Sg & 1935.921 & 1929.638 & 6.113 & 6.267 & 6.061 & 6.184& 0.284 \\
$^{263}$Db & 1933.881 & 1933.499& 6.094 & 6.270 & 6.042 & 6.180 & 0.267&& $^{264}$Sg & 1943.309 & 1937.117 & 6.120 & 6.280 & 6.067 & 6.195& 0.283 \\
$^{264}$Db & 1944.696 & 1939.319& 6.112 & 6.289 & 6.060 & 6.199 & 0.284&& $^{265}$Sg & 1948.874 & 1943.176 & 6.125 & 6.290 & 6.072 & 6.204& 0.282 \\
$^{265}$Db & 1951.545 & 1946.270& 6.113 & 6.297 & 6.060 & 6.204 & 0.273&& $^{266}$Sg & 1955.994 & 1950.424 & 6.131 & 6.303 & 6.078 & 6.214& 0.280 \\
$^{266}$Db & 1956.733 & 1952.087& 6.117 & 6.304 & 6.064 & 6.211 & 0.271&& $^{267}$Sg & 1961.514 & 1956.307 & 6.129 & 6.305 & 6.076 & 6.215& 0.270 \\
$^{267}$Db & 1962.890 & 1958.821& 6.124 & 6.319 & 6.071 & 6.223 & 0.269&& $^{268}$Sg & 1967.783 & 1963.382 & 6.136 & 6.319 & 6.083 & 6.227& 0.267 \\
$^{268}$Db & 1967.172 & 1963.905& 6.135 & 6.337 & 6.083 & 6.239 & 0.274&& $^{269}$Sg & 1972.273 & 1968.493 & 6.146 & 6.337 & 6.094 & 6.242& 0.271 \\
$^{269}$Db & 1973.558 & 1969.890& 6.138 & 6.347 & 6.086 & 6.246 & 0.267&& $^{270}$Sg & 1978.775 & 1974.836 & 6.149 & 6.346 & 6.097 & 6.249& 0.262 \\
$^{270}$Db & 1977.465 & 1974.802& 6.152 & 6.367 & 6.100 & 6.265 & 0.279&& $^{271}$Sg & 1982.783 & 1979.641 & 6.144 & 6.345 & 6.091 & 6.247& 0.235 \\
$^{271}$Db & 1983.492 & & 6.152 & 6.374 & 6.100 & 6.269 & 0.265&& $^{272}$Sg & 1989.075 & 1985.890 & 6.161 & 6.371 & 6.109 & 6.270& 0.256 \\
$^{272}$Db & 1987.240 & & 6.169 & 6.397 & 6.117 & 6.291 & 0.284&& $^{273}$Sg & 1993.036 & 1990.523 & 6.150 & 6.362 & 6.098 & 6.261& 0.213 \\
$^{273}$Db & 1993.148 & & 6.179 & 6.415 & 6.127 & 6.306 & 0.289&& $^{274}$Sg & 1998.889 &  & 6.189 & 6.412 & 6.137 & 6.307& 0.281 \\
$^{274}$Db & 1996.716 & & 6.150 & 6.385 & 6.097 & 6.276 & 0.206&& $^{275}$Sg & 2003.236 &  & 6.159 & 6.385 & 6.107 & 6.279& 0.201 \\
$^{275}$Db & 2002.428 & & 6.155 & 6.399 & 6.103 & 6.287 & 0.201&& $^{276}$Sg & 2009.100 &  & 6.166 & 6.398 & 6.114 & 6.291& 0.198 \\
$^{276}$Db & 2006.164 & & 6.148 & 6.396 & 6.096 & 6.284 & 0.166&& $^{277}$Sg & 2012.979 &  & 6.168 & 6.406 & 6.115 & 6.297& 0.188 \\
$^{277}$Db & 2011.774 & & 6.154 & 6.409 & 6.102 & 6.294 & 0.161&& $^{278}$Sg & 2018.739 &  & 6.165 & 6.409 & 6.112 & 6.298& 0.160 \\
$^{278}$Db & 2015.470 & & 6.156 & 6.417 & 6.104 & 6.301 & 0.152&& $^{279}$Sg & 2022.596 &  & 6.167 & 6.417 & 6.115 & 6.304& 0.150 \\
$^{279}$Db & 2021.156 & & 6.168 & 6.433 & 6.116 & 6.316 & -0.152&& $^{280}$Sg & 2028.235 &  & 6.179 & 6.433 & 6.127 & 6.319& -0.153 \\
$^{280}$Db & 2025.013 & & 6.171 & 6.441 & 6.119 & 6.322 & -0.146&& $^{281}$Sg & 2032.219 &  & 6.174 & 6.433 & 6.121 & 6.317& -0.120 \\
$^{281}$Db & 2030.557 & & 6.169 & 6.446 & 6.117 & 6.325 & -0.118&& $^{282}$Sg & 2037.928 &  & 6.179 & 6.446 & 6.127 & 6.328& -0.116 \\
$^{282}$Db & 2034.316 & & 6.172 & 6.454 & 6.120 & 6.332 & -0.115&& $^{283}$Sg & 2041.825 &  & 6.182 & 6.454 & 6.130 & 6.335& -0.113 \\
$^{283}$Db & 2039.536 & & 6.178 & 6.469 & 6.126 & 6.344 & -0.112&& $^{284}$Sg & 2047.186 &  & 6.188 & 6.468 &  &  & -0.111 \\
$^{284}$Db & 2042.993 & & 6.184 & 6.479 & 6.133 & 6.353 & -0.118&& $^{285}$Sg & 2050.767 &  & 6.194 & 6.479 & 6.142 & 6.356& -0.117 \\
$^{285}$Db & 2047.866 & & 6.188 & 6.493 & 6.136 & 6.364 & -0.104&& $^{286}$Sg & 2055.798 &  & 6.198 & 6.492 & 6.146 & 6.366& -0.103 \\
$^{286}$Db & 2051.354 & & 6.177 & 6.494 & 6.125 & 6.361 & -0.058&& $^{287}$Sg & 2059.205 &  & 6.187 & 6.493 & 6.135 & 6.364& 0.038 \\
$^{287}$Db & 2056.027 & & 6.183 & 6.508 & 6.131 & 6.373 & -0.046&& $^{288}$Sg & 2064.193 &  & 6.193 & 6.507 & 6.141 & 6.375& -0.045 \\
$^{288}$Db & 2059.728 & & 6.180 & 6.517 & 6.129 & 6.378 & 0.020&& $^{289}$Sg & 2068.006 &  & 6.190 & 6.515 & 6.138 & 6.380& 0.020 \\
$^{289}$Db & 2064.000 & & 6.184 & 6.530 & 6.132 & 6.389 & 0.000&& $^{290}$Sg & 2072.468 &  & 6.195 & 6.53 & 6.144 & 6.391& 0.000 \\
$^{290}$Db & 2066.274 & & 6.195 & 6.544 & 6.143 & 6.402 & -0.030&& $^{291}$Sg & 2074.896 &  & 6.204 & 6.543 & 6.153 & 6.403& -0.029 \\
$^{291}$Db & 2069.897 & & 6.215 & 6.564 & 6.164 & 6.423 & 0.061&& $^{292}$Sg & 2079.236 &  & 6.218 & 6.558 & 6.167 & 6.419& -0.001 \\

\cline{1-8}\cline{10-17}
\end{tabular}}
}
\end{table}

\begin{table}[!htbp]
\caption{Same as Table \ref{gs-properties-Md-Lo} but for Bh and Hs isotopes.}
\centering
\resizebox{1.0\textwidth}{!}{%
{\begin{tabular}{ccccccccccccccccc}

\cline{1-8}\cline{10-17}
 \multicolumn{1}{c}{Nucleus}&
 \multicolumn{2}{c}{B.E.}&
  \multicolumn{1}{c}{}&\multicolumn{1}{c}{}& \multicolumn{1}{c}{}& \multicolumn{1}{c}{}& \multicolumn{1}{c}{}&&
  \multicolumn{1}{c}{Nucleus}&
 \multicolumn{2}{c}{B.E.}&
  \multicolumn{1}{c}{}&\multicolumn{1}{c}{}& \multicolumn{1}{c}{}& \multicolumn{1}{c}{}& \multicolumn{1}{c}{}\\
   \cline{2-3} \cline{11-12}
 \multicolumn{1}{c}{}&
 \multicolumn{1}{c}{RMF}& \multicolumn{1}{c}{Expt.}&
  \multicolumn{1}{c}{$R_c$}&\multicolumn{1}{c}{$R_n$}&\multicolumn{1}{c}{$R_p$}& \multicolumn{1}{c}{$R_m$}& \multicolumn{1}{c}{$\beta$}&&
  \multicolumn{1}{c}{ }&
 \multicolumn{1}{c}{RMF}& \multicolumn{1}{c}{Expt.}&
  \multicolumn{1}{c}{$R_c$}&\multicolumn{1}{c}{$R_n$}&\multicolumn{1}{c}{$R_p$}& \multicolumn{1}{c}{$R_m$}& \multicolumn{1}{c}{$\beta$}\\
\cline{1-8}\cline{10-17}
$^{250}$Bh & 1825.891 & & 6.053 & 6.115 & 6.000 & 6.066 & 0.295&& $^{253}$Hs & 1844.275 &  & 6.059 & 6.125 & 6.006 & 6.074& 0.264 \\
$^{251}$Bh & 1836.058 & & 6.056 & 6.126 & 6.003 & 6.074 & 0.292&& $^{254}$Hs & 1854.365 &  & 6.065 & 6.138 & 6.012 & 6.085& 0.265 \\
$^{252}$Bh & 1843.691 & & 6.041 & 6.118 & 5.988 & 6.063 & 0.258&& $^{255}$Hs & 1862.001 &  & 6.069 & 6.146 & 6.016 & 6.091& 0.263 \\
$^{253}$Bh & 1853.571 & & 6.064 & 6.145 & 6.011 & 6.089 & 0.288&& $^{256}$Hs & 1871.630 &  & 6.075 & 6.160 & 6.022 & 6.102& 0.265 \\
$^{254}$Bh & 1861.042 & & 6.052 & 6.140 & 5.999 & 6.081 & 0.259&& $^{257}$Hs & 1878.995 &  & 6.093 & 6.182 & 6.040 & 6.123& 0.289 \\
$^{255}$Bh & 1870.450 & & 6.059 & 6.155 & 6.005 & 6.092 & 0.261&& $^{258}$Hs & 1888.324 &  & 6.086 & 6.182 & 6.033 & 6.120& 0.265 \\
$^{256}$Bh & 1877.700 & & 6.076 & 6.175 & 6.023 & 6.112 & 0.285&& $^{259}$Hs & 1895.386 &  & 6.091 & 6.192 & 6.038 & 6.128& 0.266 \\
$^{257}$Bh & 1886.855 & & 6.080 & 6.186 & 6.028 & 6.121 & 0.282&& $^{260}$Hs & 1904.246 &  & 6.096 & 6.204 & 6.043 & 6.138& 0.263 \\
$^{258}$Bh & 1893.736 & & 6.085 & 6.195 & 6.032 & 6.128 & 0.281&& $^{261}$Hs & 1911.104 &  & 6.101 & 6.214 & 6.049 & 6.146& 0.265 \\
$^{259}$Bh & 1902.479 & & 6.084 & 6.201 & 6.031 & 6.131 & 0.269&& $^{262}$Hs & 1919.665 &  & 6.107 & 6.226 & 6.055 & 6.156& 0.260 \\
$^{260}$Bh & 1909.209 & 1901.508& 6.089 & 6.211 & 6.036 & 6.140 & 0.269&& $^{263}$Hs & 1926.258 & 1918.585 & 6.113 & 6.237 & 6.060 & 6.165& 0.260 \\
$^{261}$Bh & 1917.586 & 1909.770& 6.096 & 6.224 & 6.043 & 6.151 & 0.266&& $^{264}$Hs & 1934.408 & 1926.771 & 6.119 & 6.250 & 6.066 & 6.176& 0.259 \\
$^{262}$Bh & 1923.963 & 1916.433& 6.101 & 6.234 & 6.049 & 6.159 & 0.265&& $^{265}$Hs & 1940.57 & 1933.505 & 6.125 & 6.260 & 6.072 & 6.184& 0.260 \\
$^{263}$Bh & 1931.919 & 1924.549& 6.108 & 6.248 & 6.055 & 6.170 & 0.264&& $^{266}$Hs & 1948.484 & 1941.341 & 6.131 & 6.274 & 6.079 & 6.195& 0.259 \\
$^{264}$Bh & 1937.982 & 1931.059& 6.119 & 6.264 & 6.066 & 6.185 & 0.272&& $^{267}$Hs & 1954.499 & 1947.895 & 6.136 & 6.283 & 6.084 & 6.203& 0.258 \\
$^{265}$Bh & 1945.576 & 1938.773& 6.120 & 6.272 & 6.067 & 6.190 & 0.262&& $^{268}$Hs & 1961.988 & 1955.790 & 6.143 & 6.297 & 6.090 & 6.215& 0.259 \\
$^{266}$Bh & 1951.322 & 1945.155& 6.124 & 6.280 & 6.072 & 6.197 & 0.260&& $^{269}$Hs & 1967.948 & 1962.127 & 6.147 & 6.306 & 6.095 & 6.222& 0.260 \\
$^{267}$Bh & 1958.580 & 1952.565& 6.146 & 6.309 & 6.093 & 6.223 & 0.285&& $^{270}$Hs & 1974.717 & 1969.650 & 6.153 & 6.319 & 6.101 & 6.233& 0.256 \\
$^{268}$Bh & 1964.571 & 1958.595& 6.138 & 6.305 & 6.085 & 6.218 & 0.267&& $^{271}$Hs & 1979.751 & 1975.094 & 6.154 & 6.324 & 6.102 & 6.237& 0.241 \\
$^{269}$Bh & 1971.068 & 1965.995& 6.150 & 6.325 & 6.098 & 6.235 & 0.272&& $^{272}$Hs & 1986.628 & 1981.899 & 6.164 & 6.343 & 6.112 & 6.252& 0.246 \\
$^{270}$Bh & 1975.615 & 1971.319& 6.145 & 6.324 & 6.093 & 6.233 & 0.246&& $^{273}$Hs & 1991.333 & 1987.086 & 6.168 & 6.351 & 6.116 & 6.259& 0.240 \\
$^{271}$Bh & 1982.445 & 1977.695& 6.165 & 6.354 & 6.113 & 6.260 & 0.272&& $^{274}$Hs & 1997.866 & 1993.561 & 6.174 & 6.364 & 6.122 & 6.269& 0.235 \\
$^{272}$Bh & 1986.671 & 1982.898& 6.157 & 6.350 & 6.105 & 6.255 & 0.242&& $^{275}$Hs & 2002.430 & 1998.499 & 6.153 & 6.344 & 6.101 & 6.249& 0.163 \\
$^{273}$Bh & 1993.047 & 1989.125& 6.166 & 6.365 & 6.113 & 6.268 & 0.242&& $^{276}$Hs & 2008.778 & 2004.905 & 6.174 & 6.374 & 6.122 & 6.277& 0.204 \\
$^{274}$Bh & 1997.097 & 1994.148& 6.169 & 6.372 & 6.117 & 6.274 & 0.234&& $^{277}$Hs & 2013.346 & 2009.768 & 6.171 & 6.377 & 6.119 & 6.277& 0.181 \\
$^{275}$Bh & 2003.389 & 2000.210& 6.178 & 6.389 & 6.126 & 6.288 & 0.238&& $^{278}$Hs & 2019.542 &  & 6.176 & 6.389 & 6.124 & 6.288& 0.176 \\
$^{276}$Bh & 2007.680 & & 6.166 & 6.382 & 6.114 & 6.279 & 0.195&& $^{279}$Hs & 2023.760 &  & 6.179 & 6.397 & 6.127 & 6.294& 0.167 \\
$^{277}$Bh & 2013.719 & & 6.166 & 6.388 & 6.114 & 6.284 & 0.176&& $^{280}$Hs & 2029.806 &  & 6.185 & 6.410 & 6.133 & 6.305& 0.163 \\
$^{278}$Bh & 2017.733 & & 6.169 & 6.397 & 6.117 & 6.291 & 0.169&& $^{281}$Hs & 2033.921 &  & 6.187 & 6.418 & 6.135 & 6.311& 0.154 \\
$^{279}$Bh & 2023.635 & & 6.175 & 6.410 & 6.123 & 6.301 & 0.166&& $^{282}$Hs & 2039.740 &  & 6.193 & 6.431 & 6.141 & 6.321& 0.150 \\
$^{280}$Bh & 2027.565 & & 6.176 & 6.417 & 6.124 & 6.307 & 0.153&& $^{283}$Hs & 2043.721 &  & 6.190 & 6.43 & 6.138 & 6.320& 0.115 \\
$^{281}$Bh & 2033.254 & & 6.182 & 6.430 & 6.130 & 6.317 & 0.149&& $^{284}$Hs & 2049.520 &  & 6.195 & 6.443 & 6.143 & 6.331& 0.111 \\
$^{282}$Bh & 2035.103 & & 6.167 & 6.410 & 6.115 & 6.300 & -0.010&& $^{285}$Hs & 2053.619 &  & 6.201 & 6.453 & 6.149 & 6.339& -0.112 \\
$^{283}$Bh & 2042.753 & & 6.185 & 6.443 & 6.133 & 6.328 & 0.110&& $^{286}$Hs & 2059.267 &  & 6.207 & 6.467 & 6.155 & 6.351& -0.110 \\
$^{284}$Bh & 2046.775 & & 6.191 & 6.453 & 6.139 & 6.337 & -0.113&& $^{287}$Hs & 2063.106 &  & 6.213 & 6.477 & 6.161 & 6.360& -0.115 \\
$^{285}$Bh & 2052.280 & & 6.198 & 6.467 & 6.146 & 6.348 & -0.112&& $^{288}$Hs & 2068.523 &  & 6.219 & 6.492 & 6.167 & 6.372& -0.106 \\
$^{286}$Bh & 2055.991 & & 6.203 & 6.477 & 6.152 & 6.357 & -0.116&& $^{289}$Hs & 2072.358 &  & 6.207 & 6.492 & 6.155 & 6.368& -0.056 \\
$^{287}$Bh & 2061.229 & & 6.210 & 6.492 & 6.158 & 6.369 & -0.107&& $^{290}$Hs & 2077.638 &  & 6.210 & 6.504 & 6.158 & 6.378& -0.001 \\
$^{288}$Bh & 2064.891 & & 6.197 & 6.492 & 6.145 & 6.365 & -0.058&& $^{291}$Hs & 2081.508 &  & 6.208 & 6.512 & 6.157 & 6.383& -0.001 \\
$^{289}$Bh & 2070.060 & & 6.200 & 6.505 & 6.149 & 6.375 & 0.000&& $^{292}$Hs & 2086.226 &  & 6.215 & 6.527 & 6.164 & 6.395& 0.000 \\
$^{290}$Bh & 2073.827 & & 6.199 & 6.513 & 6.147 & 6.381 & 0.000&& $^{293}$Hs & 2088.969 &  & 6.223 & 6.539 & 6.172 & 6.406& -0.017 \\
$^{291}$Bh & 2078.411 & & 6.205 & 6.528 & 6.153 & 6.392 & 0.000&& $^{294}$Hs & 2093.810 &  & 6.238 & 6.556 & 6.186 & 6.422& -0.001 \\
$^{292}$Bh & 2080.950 & & 6.214 & 6.540 & 6.163 & 6.404 & -0.015&&  &  &  &  &  &  & &  \\
$^{293}$Bh & 2085.663 & & 6.228 & 6.557 & 6.177 & 6.420 & 0.000&&  &  &  &  &  &  & &  \\

\cline{1-8}\cline{10-17}
\end{tabular}}
}
\end{table}

\begin{table}[!htbp]
\caption{Same as Table \ref{gs-properties-Md-Lo} but for Mt and Ds isotopes.}
\centering
\resizebox{1.0\textwidth}{!}{%
{\begin{tabular}{ccccccccccccccccc}

\cline{1-8}\cline{10-17}
 \multicolumn{1}{c}{Nucleus}&
 \multicolumn{2}{c}{B.E.}&
  \multicolumn{1}{c}{}&\multicolumn{1}{c}{}& \multicolumn{1}{c}{}& \multicolumn{1}{c}{}& \multicolumn{1}{c}{}&&
  \multicolumn{1}{c}{Nucleus}&
 \multicolumn{2}{c}{B.E.}&
  \multicolumn{1}{c}{}&\multicolumn{1}{c}{}& \multicolumn{1}{c}{}& \multicolumn{1}{c}{}& \multicolumn{1}{c}{}\\
   \cline{2-3} \cline{11-12}
 \multicolumn{1}{c}{}&
 \multicolumn{1}{c}{RMF}& \multicolumn{1}{c}{Expt.}&
  \multicolumn{1}{c}{$R_c$}&\multicolumn{1}{c}{$R_n$}&\multicolumn{1}{c}{$R_p$}& \multicolumn{1}{c}{$R_m$}& \multicolumn{1}{c}{$\beta$}&&
  \multicolumn{1}{c}{ }&
 \multicolumn{1}{c}{RMF}& \multicolumn{1}{c}{Expt.}&
  \multicolumn{1}{c}{$R_c$}&\multicolumn{1}{c}{$R_n$}&\multicolumn{1}{c}{$R_p$}& \multicolumn{1}{c}{$R_m$}& \multicolumn{1}{c}{$\beta$}\\
\cline{1-8}\cline{10-17}

$^{255}$Mt & 1853.235 & & 6.089 & 6.153 & 6.036 & 6.103 & 0.285&& $^{257}$Ds & 1861.206 &  & 6.098 & 6.157 & 6.045 & 6.109& 0.270 \\
$^{256}$Mt & 1861.001 & & 6.090 & 6.158 & 6.037 & 6.107 & 0.280&& $^{258}$Ds & 1871.198 &  & 6.104 & 6.17 & 6.052 & 6.12& 0.271 \\
$^{257}$Mt & 1870.868 & & 6.081 & 6.159 & 6.028 & 6.104 & 0.257&& $^{259}$Ds & 1878.951 &  & 6.121 & 6.191 & 6.069 & 6.14& 0.293 \\
$^{258}$Mt & 1878.472 & & 6.086 & 6.169 & 6.033 & 6.112 & 0.259&& $^{260}$Ds & 1888.664 &  & 6.113 & 6.191 & 6.060 & 6.136& 0.267 \\
$^{259}$Mt & 1887.950 & & 6.091 & 6.181 & 6.038 & 6.121 & 0.256&& $^{261}$Ds & 1896.110 &  & 6.117 & 6.200 & 6.064 & 6.143& 0.267 \\
$^{260}$Mt & 1895.106 & & 6.113 & 6.204 & 6.060 & 6.144 & 0.283&& $^{262}$Ds & 1905.355 &  & 6.121 & 6.210 & 6.068 & 6.151& 0.261 \\
$^{261}$Mt & 1904.345 & & 6.101 & 6.202 & 6.048 & 6.138 & 0.254&& $^{263}$Ds & 1912.591 &  & 6.125 & 6.220 & 6.073 & 6.159& 0.264 \\
$^{262}$Mt & 1911.389 & & 6.107 & 6.213 & 6.054 & 6.147 & 0.258&& $^{264}$Ds & 1921.523 &  & 6.129 & 6.230 & 6.077 & 6.167& 0.256 \\
$^{263}$Mt & 1920.248 & & 6.112 & 6.224 & 6.060 & 6.157 & 0.252&& $^{265}$Ds & 1928.606 &  & 6.134 & 6.241 & 6.082 & 6.175& 0.255 \\
$^{264}$Mt & 1927.056 & & 6.118 & 6.235 & 6.066 & 6.166 & 0.252&& $^{266}$Ds & 1937.163 &  & 6.139 & 6.253 & 6.087 & 6.184& 0.252 \\
$^{265}$Mt & 1935.442 & 1926.945& 6.124 & 6.248 & 6.072 & 6.176 & 0.250&& $^{267}$Ds & 1943.703 & 1935.104 & 6.144 & 6.262 & 6.092 & 6.193& 0.252 \\
$^{266}$Mt & 1941.753 & 1933.733& 6.130 & 6.258 & 6.077 & 6.185 & 0.251&& $^{268}$Ds & 1952.009 & 1943.407 & 6.149 & 6.275 & 6.097 & 6.202& 0.249 \\
$^{267}$Mt & 1949.870 & 1941.975& 6.136 & 6.271 & 6.083 & 6.195 & 0.249&& $^{269}$Ds & 1958.442 & 1950.291 & 6.154 & 6.284 & 6.102 & 6.210& 0.249 \\
$^{268}$Mt & 1956.077 & 1948.686& 6.141 & 6.281 & 6.089 & 6.204 & 0.249&& $^{270}$Ds & 1966.276 & 1958.519 & 6.159 & 6.297 & 6.107 & 6.220& 0.247 \\
$^{269}$Mt & 1963.687 & 1956.539& 6.147 & 6.294 & 6.095 & 6.214 & 0.248&& $^{271}$Ds & 1972.587 & 1965.321 & 6.165 & 6.308 & 6.113 & 6.229& 0.251 \\
$^{270}$Mt & 1969.833 & 1963.270& 6.153 & 6.305 & 6.101 & 6.224 & 0.255&& $^{272}$Ds & 1979.993 & 1973.325 & 6.170 & 6.319 & 6.117 & 6.238& 0.243 \\
$^{271}$Mt & 1976.857 & 1970.950& 6.157 & 6.316 & 6.105 & 6.232 & 0.241&& $^{273}$Ds & 1985.570 & 1979.055 & 6.165 & 6.319 & 6.113 & 6.237& 0.222 \\
$^{272}$Mt & 1982.298 & 1976.541& 6.159 & 6.322 & 6.107 & 6.237 & 0.233&& $^{274}$Ds & 1992.845 & 1986.286 & 6.171 & 6.331 & 6.119 & 6.247& 0.217 \\
$^{273}$Mt & 1989.347 & 1983.481& 6.160 & 6.329 & 6.107 & 6.241 & 0.218&& $^{275}$Ds & 1998.061 & 1991.989 & 6.174 & 6.338 & 6.122 & 6.252& 0.209 \\
$^{274}$Mt & 1994.284 & 1989.019& 6.163 & 6.335 & 6.111 & 6.247 & 0.212&& $^{276}$Ds & 2005.142 & 1999.084 & 6.181 & 6.351 & 6.129 & 6.264& 0.206 \\
$^{275}$Mt & 2001.156 & 1995.507& 6.170 & 6.351 & 6.118 & 6.260 & 0.211&& $^{277}$Ds & 2010.196 & 2004.557 & 6.184 & 6.359 & 6.132 & 6.270& 0.200 \\
$^{276}$Mt & 2005.950 & 2001.093& 6.175 & 6.359 & 6.122 & 6.267 & 0.205&& $^{278}$Ds & 2016.903 & 2011.387 & 6.191 & 6.373 & 6.139 & 6.281& 0.197 \\
$^{277}$Mt & 2012.456 & 2007.511& 6.181 & 6.374 & 6.129 & 6.279 & 0.203&& $^{279}$Ds & 2021.904 & 2016.713 & 6.189 & 6.377 & 6.137 & 6.284& 0.179 \\
$^{278}$Mt & 2017.141 & 2012.814& 6.185 & 6.384 & 6.133 & 6.287 & 0.198&& $^{280}$Ds & 2028.422 & 2023.391 & 6.195 & 6.390 & 6.143 & 6.294& 0.175 \\
$^{279}$Mt & 2023.486 & 2019.126& 6.193 & 6.398 & 6.141 & 6.299 & 0.197&& $^{281}$Ds & 2033.076 & 2028.551 & 6.198 & 6.399 & 6.147 & 6.301& 0.168 \\
$^{280}$Mt & 2027.788 & & 6.196 & 6.407 & 6.144 & 6.306 & 0.19&&   $^{282}$Ds & 2039.358 &  & 6.204 & 6.411 & 6.152 & 6.311& 0.164 \\
$^{281}$Mt & 2033.929 & & 6.193 & 6.409 & 6.141 & 6.306 & 0.162&&  $^{283}$Ds & 2043.779 &  & 6.207 & 6.420 & 6.155 & 6.319& 0.158 \\
$^{282}$Mt & 2038.193 & & 6.196 & 6.417 & 6.144 & 6.313 & 0.155&&  $^{284}$Ds & 2049.868 &  & 6.213 & 6.432 & 6.161 & 6.328& 0.152 \\
$^{283}$Mt & 2044.160 & & 6.202 & 6.430 & 6.150 & 6.324 & 0.152&&  $^{285}$Ds & 2054.024 &  & 6.212 & 6.436 & 6.161 & 6.331& 0.138 \\
$^{284}$Mt & 2048.156 & & 6.202 & 6.436 & 6.150 & 6.328 & 0.139&&  $^{286}$Ds & 2060.049 &  & 6.222 & 6.453 & 6.171 & 6.346& 0.141 \\
$^{285}$Mt & 2054.045 & & 6.212 & 6.452 & 6.160 & 6.342 & 0.142&&  $^{287}$Ds & 2064.281 &  & 6.216 & 6.450 & 6.164 & 6.342& 0.104\\
$^{286}$Mt & 2058.013 & & 6.207 & 6.450 & 6.156 & 6.340 & 0.105&&  $^{288}$Ds & 2070.135 &  & 6.226 & 6.466 & 6.174 & 6.356& -0.108\\
$^{287}$Mt & 2063.821 & & 6.216 & 6.466 & 6.165 & 6.353 & -0.110&& $^{289}$Ds & 2074.229 &  & 6.220 & 6.469 & 6.169 & 6.356& -0.066\\
$^{288}$Mt & 2067.786 & & 6.221 & 6.476 & 6.170 & 6.362 & -0.114&& $^{290}$Ds & 2079.913 &  & 6.225 & 6.482 & 6.174 & 6.367& -0.057\\
$^{289}$Mt & 2073.29  & & 6.229 & 6.491 & 6.177 & 6.374 & -0.108&& $^{291}$Ds & 2083.96  &  & 6.226 & 6.490 & 6.175 & 6.373& -0.055 \\
$^{290}$Mt & 2077.171 & & 6.216 & 6.491 & 6.165 & 6.370 & -0.059&& $^{292}$Ds & 2089.55  &  & 6.213 & 6.488 & 6.161 & 6.367& -0.001 \\
$^{291}$Mt & 2082.58  & & 6.220 & 6.503 & 6.168 & 6.379 & -0.004&& $^{293}$Ds & 2093.601 &  & 6.227 & 6.510 & 6.176 & 6.387& -0.001 \\
$^{292}$Mt & 2086.517 & & 6.218 & 6.511 & 6.166 & 6.384 & -0.003&& $^{294}$Ds & 2098.611 &  & 6.236 & 6.525 & 6.184 & 6.400&  0.000\\
$^{293}$Mt & 2091.379 & & 6.227 & 6.526 & 6.175 & 6.397 &  0.000&& $^{295}$Ds & 2101.709 &  & 6.244 & 6.537 & 6.192 & 6.411& -0.016 \\
$^{294}$Mt & 2094.359 & & 6.234 & 6.538 & 6.183 & 6.408 & -0.018&& $^{296}$Ds & 2106.882 &  & 6.256 & 6.553 & 6.205 & 6.426& -0.001\\
$^{295}$Mt & 2099.345 & & 6.247 & 6.554 & 6.196 & 6.424 & -0.004&&   &   &  &   &   &   &  &   \\
\cline{1-8}\cline{10-17}
\end{tabular}}}
\label{gs-properties-Mt}
\end{table}

\end{document}